\documentclass[useAMS,usenatbib]{mn2e}
\usepackage{float,graphics,epsfig,array,amsmath,amssymb,times}
\usepackage{xspace,url,longtable,natbib}
\usepackage[light]{draftcopy}
\pdfoutput=1

\newcommand{\be}{\begin{equation}}  \newcommand{\ee}{\end{equation}}
  \newcommand{\ba}{\begin{eqnarray}}

\newcommand{\automask}{\texttt{automask}\xspace}
\newcommand{\BPZ}{\texttt{BPZ}\xspace}
\newcommand{\CADC}{{CADC}\xspace}
\newcommand{\CARS}{{CARS}\xspace}
\newcommand{\CFHT}{{CFHT}\xspace}

\newcommand{\CFHTLenS}{{CFHTLenS}\xspace}
\newcommand{\CFHTLS}{{CFHTLS}\xspace}
\newcommand{\Elixir}{\texttt{Elixir}\xspace}
\newcommand{\EyE}{\texttt{EyE}\xspace}
\newcommand{\LePhare}{\texttt{LePhare}\xspace}
\newcommand{\THELI}{\texttt{THELI}\xspace}
\newcommand{\SExtractor}{\texttt{SExtractor}\xspace}
\newcommand{\swarp}{\texttt{swarp}\xspace}
\newcommand{\scamp}{\texttt{scamp}\xspace}
\newcommand{\DEEPT}{{DEEP2}\xspace}
\newcommand{\MegaCam}{{MegaCam}\xspace}
\newcommand{\MegaPipe}{{MegaPipe}\xspace}
\newcommand{\MegaPrime}{{MegaPrime}\xspace}
\newcommand{\Sloan}{{SDSS}\xspace}
\newcommand{\SDSSRS}{{SDSS-DR7}\xspace}
\newcommand{\SDSSRE}{{SDSS-DR8}\xspace}

\newcommand{\Terapix}{{Terapix}\xspace}
\newcommand{\TWOMASS}{{2MASS}\xspace}
\newcommand{\VIMOS}{{VIMOS}\xspace}
\newcommand{\VIPERS}{{VIPERS}\xspace}
\newcommand{\VVDS}{{VVDS}\xspace}
\newcommand{\Wone}{{W1}\xspace}
\newcommand{\Wtwo}{{W2}\xspace}
\newcommand{\Wthree}{{W3}\xspace}
\newcommand{\Wfour}{{W4}\xspace}
\newcommand{\lensfit}{\texttt{lensfit}\xspace}
\newcommand{\myarcsec}{\hbox{$.\!\!^{\prime\prime}$}}
\newcommand{\myarcmin}{\hbox{$.\!\!^{\prime}$}}
\newcommand{\sectionref}[1]{Sect.~\ref{#1}}
\newcommand{\appendixref}[1]{Appendix~\ref{#1}}
\newcommand{\figref}[1]{Fig.~\ref{#1}}

\newcommand{\tabref}[1]{Table~\ref{#1}}

\def\ga{\mathrel{\mathchoice {\vcenter{\offinterlineskip\halign{\hfil
$\displaystyle##$\hfil\cr>\cr\sim\cr}}}
{\vcenter{\offinterlineskip\halign{\hfil$\textstyle##$\hfil\cr
>\cr\sim\cr}}}
{\vcenter{\offinterlineskip\halign{\hfil$\scriptstyle##$\hfil\cr
>\cr\sim\cr}}}
{\vcenter{\offinterlineskip\halign{\hfil$\scriptscriptstyle##$\hfil\cr
>\cr\sim\cr}}}}}
\def\la{\mathrel{\mathchoice {\vcenter{\offinterlineskip\halign{\hfil
$\displaystyle##$\hfil\cr<\cr\sim\cr}}}
{\vcenter{\offinterlineskip\halign{\hfil$\textstyle##$\hfil\cr
<\cr\sim\cr}}}
{\vcenter{\offinterlineskip\halign{\hfil$\scriptstyle##$\hfil\cr
<\cr\sim\cr}}}
{\vcenter{\offinterlineskip\halign{\hfil$\scriptscriptstyle##$\hfil\cr
<\cr\sim\cr}}}}}

\title[CFHTLenS: Imaging and Catalogue Products]
{CFHTLenS: The Canada-France-Hawaii Telescope Lensing Survey - Imaging Data and Catalogue Products}
\author[T.~Erben et~al.]
{T.~Erben$^1$,
H.~Hildebrandt$^{1,2}$, 
L.~Miller$^3$, 
L.~van Waerbeke$^2$,
C.~Heymans$^4$, 
H.~Hoekstra$^{5,6}$, \newauthor
T.D.~Kitching$^4$,
Y.~Mellier$^7$, 
J.~Benjamin$^2$,
C.~Blake$^{8}$,
C.~Bonnett$^{9}$,
O.~Cordes$^{1}$,
J.~Coupon$^{10}$, \newauthor
L.~Fu$^{11}$,
R.~Gavazzi$^{7}$,
B.~Gillis$^{12,13}$,
E.~Grocutt$^{4}$,
S.D.J.~Gwyn$^{14}$,
K.~Holhjem$^{15,1}$, \newauthor
M.J.~Hudson$^{12,13}$,
M.~Kilbinger$^{16,17,14,7}$,
K.~Kuijken$^{5}$,
M.~Milkeraitis$^{2}$,
B.T.P.~Rowe$^{19,20}$, \newauthor
T.~Schrabback$^{1,21,5}$,
E.~Semboloni$^{5}$,
P.~Simon$^{1}$,
M.~Smit$^{5}$,
O.~Toader$^{2}$,
S.~Vafaei$^{2}$,\newauthor
E.~van Uitert$^{5,1}$,
and
M.~Velander$^{3,5}.$
\\
$^1$Argelander Institute for Astronomy, University of Bonn, Auf dem H{\"u}gel 71, 53121 Bonn, Germany.\\
$^2$Department of Physics and Astronomy, University of British Columbia, 6224 Agricultural Road, Vancouver, V6T 1Z1, BC, Canada.\\
$^3$Department of Physics, Oxford University, Keble Road, Oxford OX1 3RH, UK.\\ 
$^4$Scottish Universities Physics Alliance, Institute for Astronomy, University of Edinburgh, Royal Observatory, Blackford Hill, Edinburgh, EH9 3HJ, UK. \\ 
$^5$Leiden Observatory, Leiden University, Niels Bohrweg 2, 2333 CA Leiden, The Netherlands.\\
$^6$Department of Physics and Astronomy, University of Victoria, Victoria, BC V8P 5C2, Canada.\\
$^7$Institut d'Astrophysique de Paris, UniversitÃ© Pierre et Marie Curie - Paris 6, 98 bis Boulevard Arago, F-75014 Paris, France.\\
$^{8}$Centre for Astrophysics \& Supercomputing, Swinburne University of
Technology, P.O. Box 218, Hawthorn, VIC 3122, Australia.\\
$^{9}$Institut de Ciencies de l’Espai, CSIC/IEEC, F. de Ciencies, Torre C5 par-2, Barcelona 08193, Spain.\\
$^{10}$Institute of Astronomy and Astrophysics, Academia Sinica, P.O. Box 23-141, Taipei 10617, Taiwan.\\
$^{11}$Key Lab for Astrophysics, Shanghai Normal University, 100 Guilin Road, 200234, Shanghai, China. \\
$^{12}$Department of Physics and Astronomy, University of Waterloo, Waterloo, ON, N2L 3G1, Canada.\\
$^{13}$Perimeter Institute for Theoretical Physics, 31 Caroline Street N, Waterloo, ON, N2L 1Y5, Canada.\\
$^{14}$Canadian Astronomical Data Centre, Herzberg Institute of Astrophysics
Victoria, BC, Canada.\\
$^{15}$Southern Astrophysical Research Telescope, Casilla 603, La Serena, Chile.\\
$^{16}$CEA Saclay, Service d'Astrophysique (SAp), Orme des Merisiers, B\^at 709, F-91191 Gif-sur-Yvette, France.\\
$^{17}$Excellence Cluster Universe, Boltzmannstr. 2, D-85748 Garching, Germany.\\
$^{18}$Universit\"ats-Sternwarte, Ludwig-Maximillians-Universit\"at M\"unchen, Scheinerstr.~1, 81679 M\"unchen, Germany.\\
$^{19}$Department of Physics and Astronomy, University College London, Gower Street, London WC1E 6BT, U.K.\\
$^{20}$California Institute of Technology, 1200 E California Boulevard, Pasadena CA 91125, USA.\\
$^{21}$Kavli Institute for Particle Astrophysics and Cosmology, Stanford University, 382 Via Pueblo Mall, Stanford, CA 94305-4060, USA.\\
}

\begin{document}

\pagerange{\pageref{firstpage}--\pageref{lastpage}} \pubyear{2012}

\maketitle

\label{firstpage}

\begin{abstract}
  We present data products from the Canada-France-Hawaii Telescope
  Lensing Survey (\CFHTLenS). \CFHTLenS is based on the Wide component
  of the Canada-France-Hawaii Telescope Legacy Survey (\CFHTLS). It
  encompasses 154 deg$^2$ of deep, optical, high-quality,
  sub-arcsecond imaging data in the five optical filters
  $u^*g'r'i'z'$. The scientific aims of the \CFHTLenS team are weak
  gravitational lensing studies supported by photometric redshift
  estimates for the galaxies. The article presents our data processing
  of the complete \CFHTLenS data set. We were able to obtain a data
  set with very good image quality and high-quality astrometric and
  photometric calibration. Our external astrometric accuracy is
  between 60-70 mas with respect to \Sloan data and the internal
  alignment in all filters is around 30 mas.  Our average photometric
  calibration shows a dispersion on the order of 0.01 to 0.03 mag for
  $g'r'i'z'$ and about 0.04 mag for $u^*$ with respect to \Sloan
  sources down to $i_{\rm \Sloan} \leq 21$. We demonstrate in
  accompanying articles that our data meet necessary requirements to
  fully exploit the survey for weak gravitational lensing analyses in
  connection with photometric redshift studies.  In the spirit of the
  \CFHTLS all our data products are released to the astronomical
  community via the Canadian Astronomy Data Centre at
  \url{http://www.cadc-ccda.hia-iha.nrc-cnrc.gc.ca/community/CFHTLens/query.html}. We
  give a description and \emph{how-to} manuals of the public products
  which include image pixel data, source catalogues with photometric
  redshift estimates and all relevant quantities to perform weak
  lensing studies.
\end{abstract}

\section{Introduction}
Our knowledge of the nature and the composition of the Universe has
evolved tremendously during the past decade. A combination of
observations has led to the conclusion that the Universe is dominated
by a uniformly distributed form of \emph{dark energy}. Chief evidences
for this conclusion are that the expansion rate is accelerating (from
the distances to supernovae; see e.g.~\citealp{rfc98,pag99,rsc07}),
that the Universe is flat (from the Cosmic Microwave Background; see
e.g.~\citealp{ksd11}) and that \emph{dark matter} cannot provide the
critical density (for instance through galaxy cluster studies; see
e.g. \citealp{aem11}).  As the standard accelerating Universe is set
on such solid grounds one of the main goals of cosmology is now to get
a precise understanding on the nature of dark matter and dark energy.

Complementary to the observations mentioned above, weak gravitational
lensing has been recognised as one of the most important tools to
study the invisible Universe.  Inhomogeneities in the mass
distribution cause the light coming from distant galaxies to be
deflected which leads to a direct observable distortion of galaxy
images.  Because the lensing effect is insensitive to the dynamical
and physical state of the mass constituents, surveying coherent image
distortions over large portions of the sky provides the most direct
mapping of the large scale structure in our Universe. After the first
significant measurement of this \emph{cosmic shear effect} by several
groups in a few square degrees of sky
\citep[see][]{wme00,wtk00,bre00,kwg00}, large efforts have been
undertaken to increase the sky coverage \citep[see
e.g.][]{wmr01,hyg02,jbf03,hss07,bhs07} and to improve the accuracy of
the necessary analysis techniques \citep[see
e.g.][]{ewb01,brc00,hvb06,mhb07,bsa09,kbb12,krh12,krg12}. In order to
obtain the best possible precision on galaxy shapes, the first major
requirement for shear measurement is image quality.  Current weak
lensing surveys are typically trying to measure galaxy shapes with a
goal of residual systematics of the order of one percent of the cosmic
shear signal \citep[][]{hvm12}. The second major requirement is depth
and multi-colour coverage so that photometric redshifts are reliable
for the interpretation of the lensing signal \citep[][]{hek12}. An
important aspect combining image quality and survey depth is the
number density of source galaxies for which shapes and photometric
redshifts meet the requirements.
In this article we present the
\CFHTLenS\footnote{\url{http://www.cfhtlens.org/}} data set which was
carefully designed as a weak lensing survey within the \CFHTLS. It spans 154
deg$^2$ in the five optical \Sloan-like filters
$u^*g'r'i'z'$. The survey was observed under the acronym \CFHTLS-Wide
and all data were obtained within superb observing conditions on the
Canada-France-Hawaii Telescope (\CFHT). Important cosmic shear results were
already obtained on significant parts of the survey
\citep[see][]{hmv06,smv06,fsh08,kbg09,tsu09}. However, these early results
were based on the analysis of a single passband only.

During the later stages of \CFHTLS-Wide observations, the \CFHTLenS
team was formed to combine this unique data set with the
expertise of the team in the technical fields of data processing,
shear analysis and photometric redshifts, as well as expertise to
optimally exploit lensing and photometric redshift catalogues. The
\CFHTLenS data an\-aly\-sis effort is complemented by comprehensive
simulations \citep[][]{hvw12} to evaluate shear measurement algorithms
and error estimates for cosmic shear analyses.

This article focuses on the presentation of the \CFHTLenS data set and
all the steps necessary to obtain the products required for weak
lensing experiments. A comprehensive evaluation of how well our data
products meet weak lensing requirements is given in the accompanying
\CFHTLenS articles \citet{hvm12}, \citet{mhk12} and
\citet{hek12}. This paper also describes the data products being
publicly released to the astronomical community.

The paper is organised as follows: We give a short overview of the
\CFHTLenS data set in \sectionref{sec:cfhtlens}. Our lensing
specialised data processing leading from \Elixir preprocessed
exposures to co-added imaging products is detailed
in \sectionref{sec:dataprocessing}.  Sections~\ref{sec:cosmicrays} and
\ref{sec:astromphotomqual} summarise important astrometric and
photometric quality characteristics of our data.  A short summary
on the released \CFHTLenS data products and our conclusions wind up
the main article. In the appendices we give detailed quality information
on each individual \CFHTLenS pointing (\appendixref{app:cfhtlensqual})
and provide \emph{how-to} manuals for the public \CFHTLenS imaging
and catalogue products (Appendices~\ref{sec:cfhtlensimaging} and
\ref{sec:cfhtlenscatalogues}).
\section{The Canada-France Hawaii Telescope Lensing Survey Data Set}
\label{sec:cfhtlens}
\begin{figure*}[ht]
  \centering
  \includegraphics[width=0.95\columnwidth]{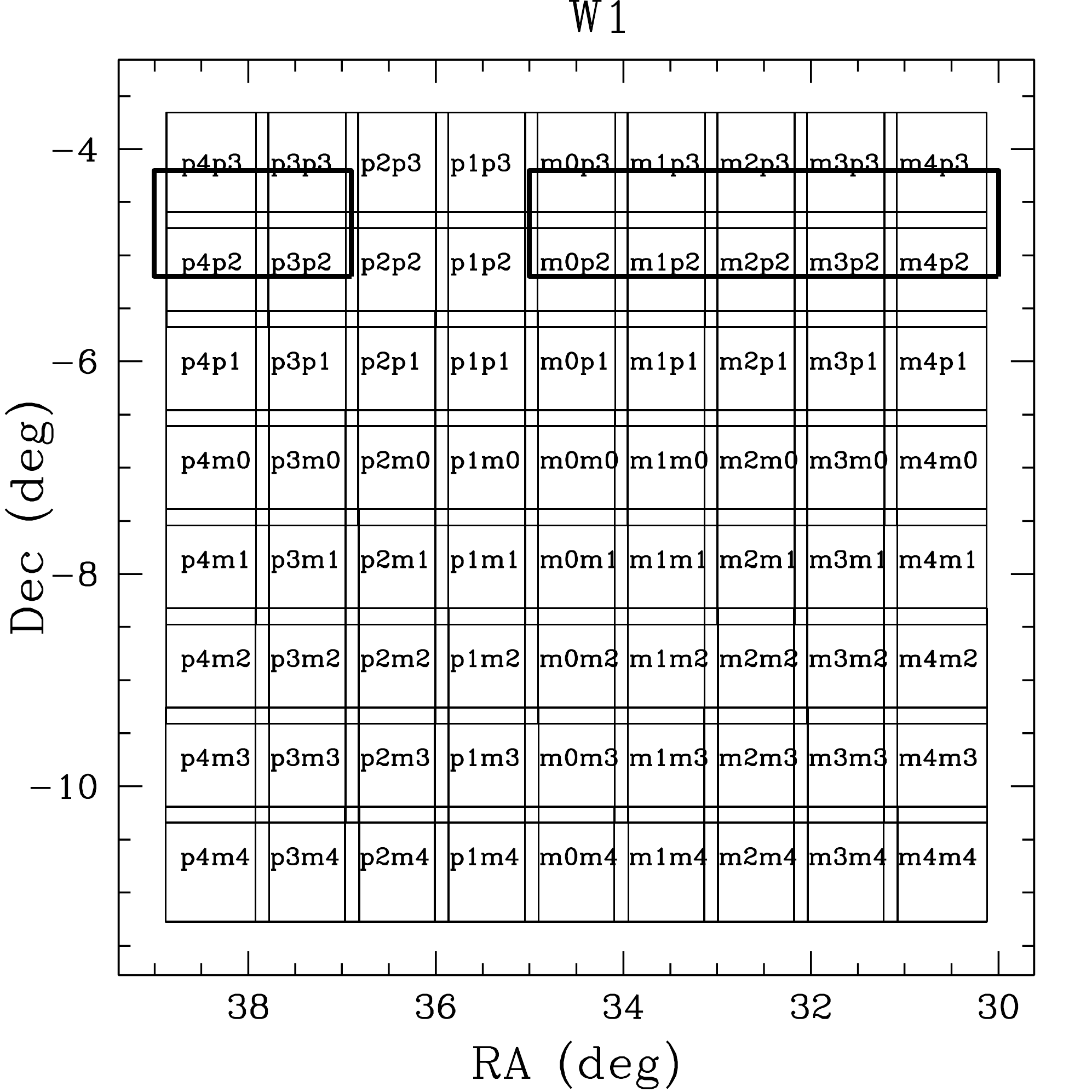}
  \includegraphics[width=0.95\columnwidth]{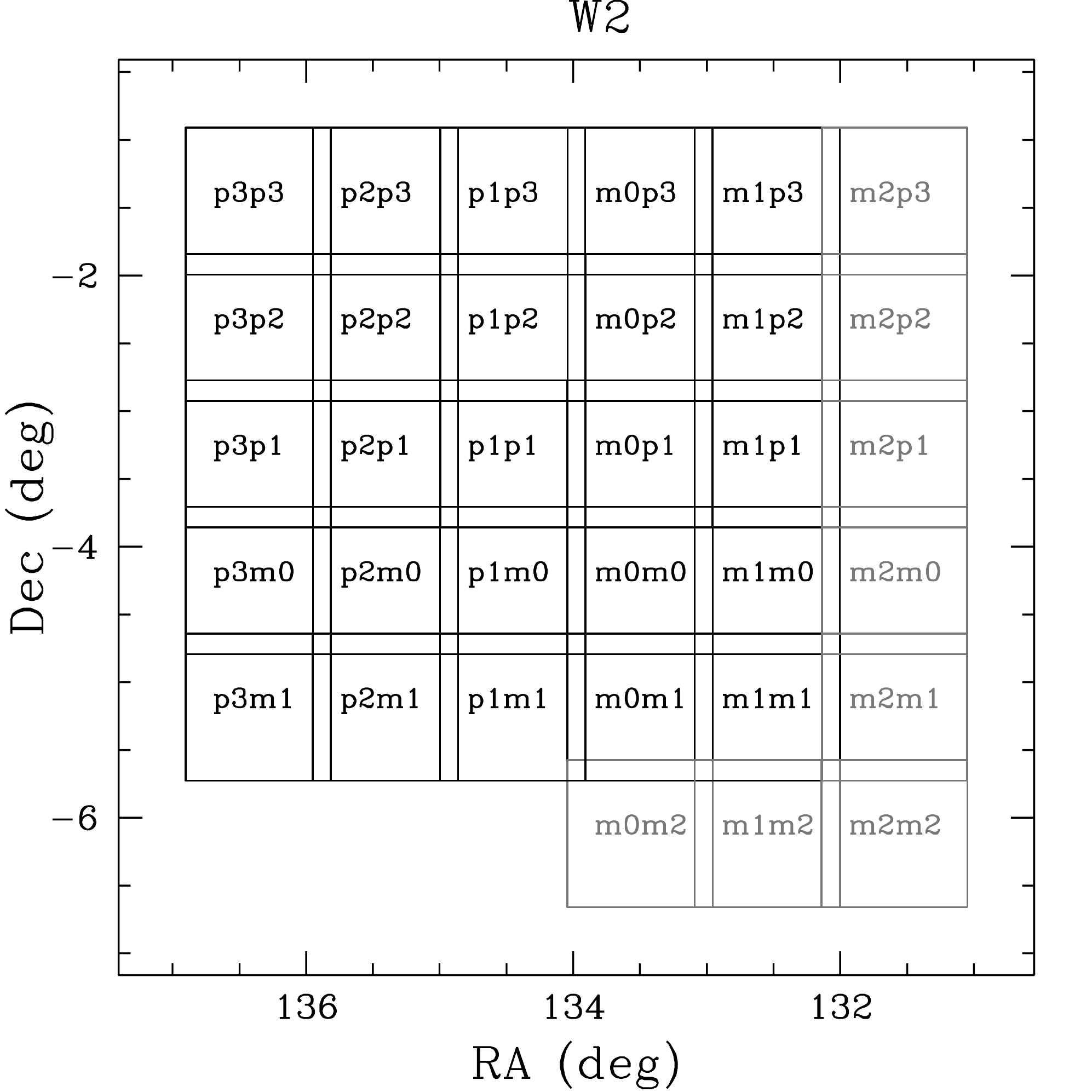}
  \includegraphics[width=0.95\columnwidth]{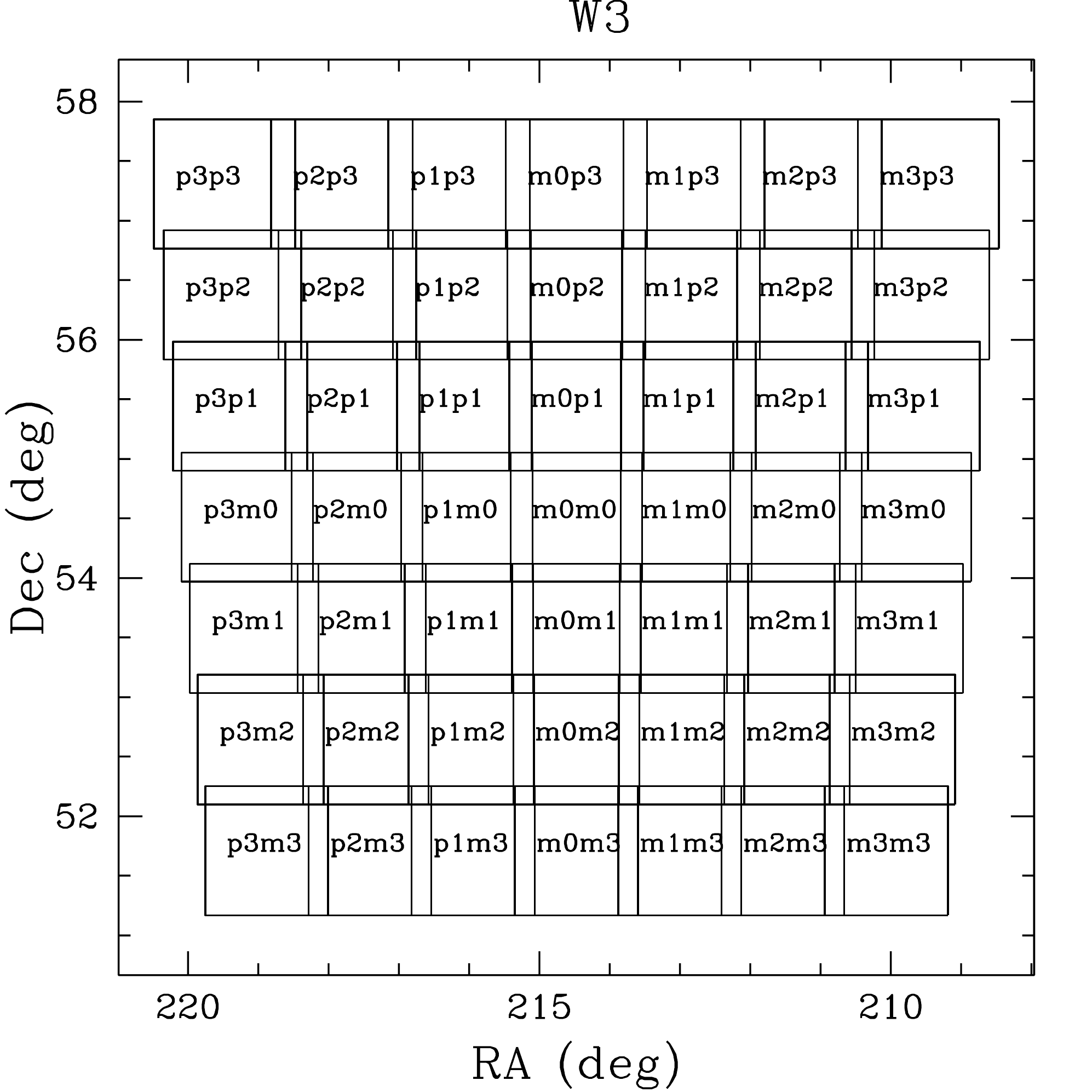}
  \includegraphics[width=0.95\columnwidth]{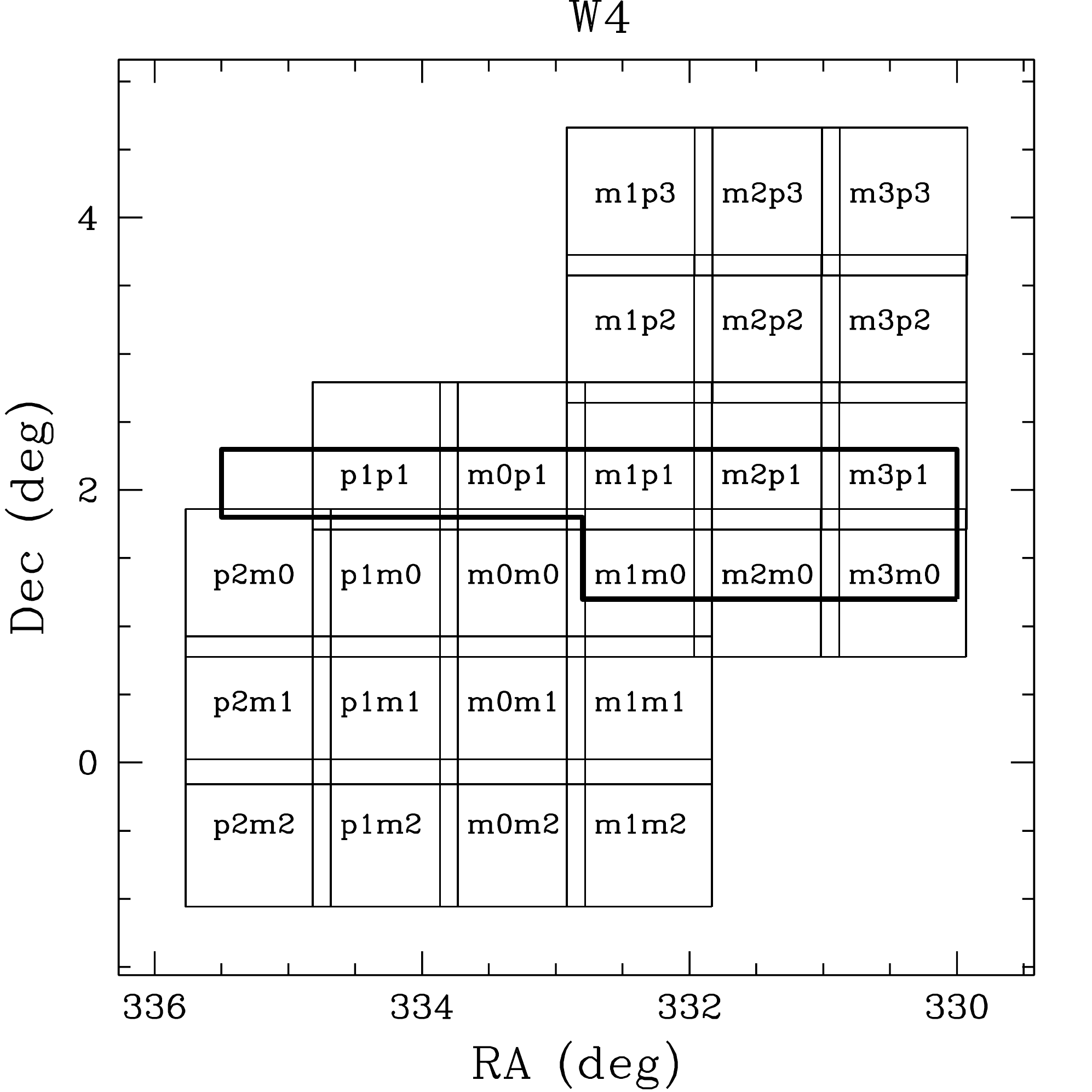}
  \caption{\label{fig:cfhtlenslayout} Layout of the four \CFHTLenS patches.
  The gray pointings in the \Wtwo region denote fields with incomplete
  colour coverage. They are not included in the \CFHTLenS project. Enclosed
  areas in \Wone and \Wfour indicate regions of available spectroscopic
  redshifts for a photometry crosscheck as discussed in
  \sectionref{sec:specz_comp}. See text for further details.}
\end{figure*}
The \CFHTLenS data set is based on the Wide part of the \CFHTLS, which
was observed in the period between 22nd of March 2003 and 1st of
November 2008. All the data were obtained with the \MegaPrime
instrument\footnote{\url{http://www.cfht.hawaii.edu/Instruments/Imaging/Megacam/}}
\citep[see][]{bca03} which is mounted on the \CFHT.  \MegaPrime is an
optical multi-chip instrument with a $9\times 4$ CCD array
($2048\times 4096$ pixels in each CCD; $0\myarcsec 187$ pixel scale;
$\sim 1^{\circ}\times 1^{\circ}$ total field-of-view).  \CFHTLS-Wide
observations were carried out in four high-galactic latitude patches:
patch \Wone with 72 pointings around RA=02$^{\rm h}$18$^{\rm
  m}$00$^{\rm s}$, Dec=$-$07$^{\rm d}$00$^{\rm m}$00$^{\rm s}$, patch
\Wtwo with 33 pointings around RA=08$^{\rm h}$54$^{\rm m}$00$^{\rm
  s}$, Dec=$-$04$^{\rm d}$15$^{\rm m}$00$^{\rm s}$, patch \Wthree with
49 pointings around RA=14$^{\rm h}$17$^{\rm m}$54$^{\rm s}$,
Dec=$+$54$^{\rm d}$30$^{\rm m}$31$^{\rm s}$ and patch \Wfour with 25
pointings around RA=22$^{\rm h}$13$^{\rm m}$18$^{\rm s}$,
Dec=$+$01$^{\rm d}$19$^{\rm m}$00$^{\rm s}$.  \CFHTLenS uses all
\CFHTLS-Wide pointings with complete colour coverage in the five
filters $u^*g'r'i'z'$. This set comprises 171 pointings with an
effective survey area of about 154 deg$^2$. The \CFHTLS-Wide patch
\Wtwo has eight additional pointings with incomplete colour coverage.
These are not included in \CFHTLenS.  The \CFHTLenS survey
layout is shown in \figref{fig:cfhtlenslayout}. Pointings are labelled
as W1m1p2 (read ``{W1} minus 1 plus 2''; see also
\figref{fig:cfhtlenslayout}).  They indicate the patch and the
separation (approximately in degrees) from the patch centre. For
instance, pointing W1m1p2 is about one degree west and two degrees
north of the W1 centre.  The overlap of adjacent pointings is about
$3\myarcmin 0$ in right ascension and $6\myarcmin 0$ in declination.

\begin{figure}
  \centering
  \includegraphics[width=0.95\columnwidth]{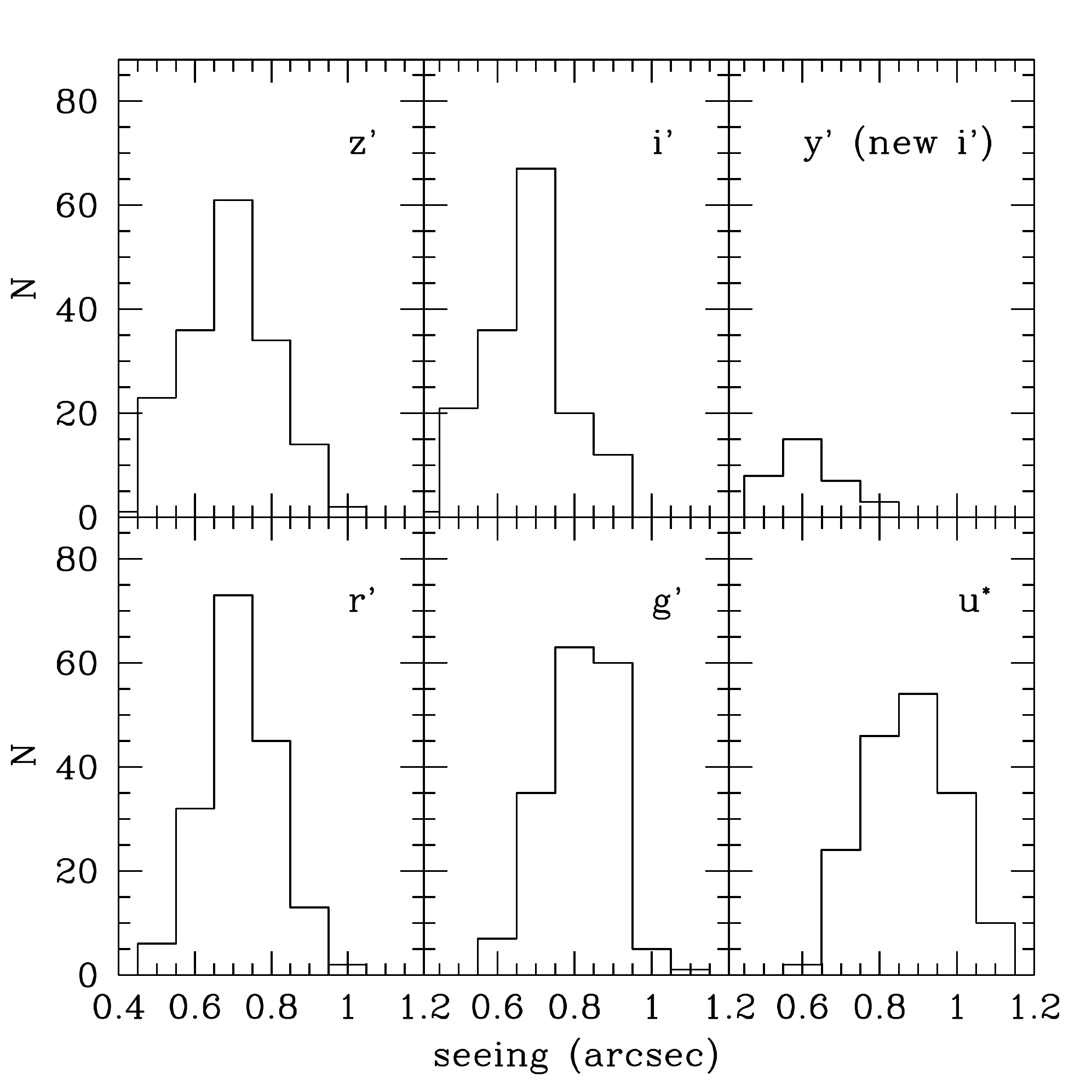}
  \caption{\label{fig:cfhtlensseeing}
    Seeing distributions for all \CFHTLenS fields and filters.}
\end{figure}

\tabref{tab:averagequal} contains observational details and provides
average quality characteristics of our co-added \CFHTLenS
pointings. It lists the targeted observing time for the different
filters, the mean limiting magnitudes and the mean seeing values with
their corresponding standard deviations over all \CFHTLenS pointings.
The seeing is estimated using the \SExtractor \citep[see][]{bea96}
parameter FWHM\_IMAGE for stellar sources. Our limiting magnitude,
$m_{\rm lim}$, is the 5-$\sigma$ detection limit in a $2\myarcsec 0$
aperture\footnote{ $m_{\rm lim}=ZP-2.5\log(5\sqrt{N_{\rm
      pix}}\sigma_{\rm sky})$, where $ZP$ is the magnitude zeropoint,
  $N_{\rm pix}$ is the number of pixels in a circle with radius
  $2\myarcsec 0$ and $\sigma_{\rm sky}$ the sky background noise
  variation.}. Nearly all 171 pointings in all filters were obtained
under superb, photometrically homogeneous and sub-arcsecond seeing
conditions (see also \tabref{tab:CFHTLenSquality}). In
\figref{fig:cfhtlensseeing} we show the full seeing distribution for
all fields and filters. It does not show the skewness to large values
that is typical in large and long-term observing campaigns without
imposed seeing constraints.

We note that the original \CFHT $i'$-band filter (\CFHT
identification: i.MP9701) broke in 2008 and a total of 33 fields were
obtained with its successor (\CFHT identification: i.MP9702).  19
fields, whose PSF properties in the original $i'$-band observations
were classified as problematic for weak lensing studies, have
observations in both filters. If necessary we distinguish the two
with labels $i'$ for i.MP9701 and $y'$ for i.MP9702.  A table
detailing important quality properties for each pointing and filter is
given in \appendixref{app:cfhtlensqual}.
\begin{table}
  \caption{Characteristics of the final \CFHTLenS co-added science data
    (see the text for an explanation of the columns).}           
\label{tab:averagequal}      
\centering          
\begin{tabular}{llll}    
\hline\hline       
\multicolumn{1}{c}{Filter} & \multicolumn{1}{c}{expos. time [s]} &
\multicolumn{1}{c}{$m_{\rm lim}$ [AB mag]} & \multicolumn{1}{c}{seeing [$''$]}\\
& & \multicolumn{1}{c}{5-$\sigma$ lim. mag.} & \\
& & \multicolumn{1}{c}{in a $2\myarcsec 0$ aperture} & \\
\hline
$u^* (u.MP9301) $ & $5\times 600$ (3000) & $25.24 \pm 0.17$ & $0.88 \pm 0.11$ \\ 
$g' (g.MP9401) $ & $5\times 500$ (2500) & $25.58 \pm 0.15$ & $0.82 \pm 0.10$ \\ 
$r' (r.MP9601) $ & $4\times 500$ (2000) & $24.88 \pm 0.16$ & $0.72 \pm 0.09$ \\ 
$i' (i.MP9701) $ & $7\times 615$ (4305) & $24.54 \pm 0.19$ & $0.68 \pm 0.11$ \\ 
$y' (i.MP9702) $ & $7\times 615$ (4305) & $24.71 \pm 0.13$ & $0.62 \pm 0.09$ \\ 
$z' (z.MP9801) $ & $6\times 600$ (3600) & $23.46 \pm 0.20$ & $0.70 \pm 0.12$ \\ 
\hline                  
\end{tabular}
\end{table}
\section{Data processing}
\label{sec:dataprocessing}
The primary goal of the image processing modules we created is to
provide the following products, necessary for the weak lensing and
photometric redshift analyses.
\begin{enumerate}
\item Deep, co-added astrometrically and photometrically calibrated
  images for all \CFHTLenS pointings in each filter.
  These images are primarily used to define the source
  catalogue sample for our lensing studies and to estimate photometric
  redshifts; see \citet{hek12}. A short summary can be found
  in \appendixref{sec:cfhtlenscatalogues}. Each co-added science image
  is accompanied by an inverse-variance \emph{weight} map which
  describes its noise
  properties \citep[see, e.g., Fig. 2 of][]{ehl09}. In addition, we
  create a so-called \emph{sum} image. This is an
  integer-value image which gives, for each pixel of the co-added
  science image, the number of single frames that contribute to that
  pixel. It is used to easily identify image regions that do not reach
  the full survey depth, such as areas around chip or edge boundaries.
\item For the $i'$-filter observations, which are used for our shape
  and lensing analysis, we require sky-subtracted individual chips that
  are not co-added. They are accompanied by bad-pixel maps, cosmic
  ray masks, and precise information of astrometric distortions and
  photometric properties. In connection with the object catalogues
  extracted from the co-added images, these products are primarily
  used by our \lensfit weak shear measurement pipeline. The procedures
  to model the PSF and to determine object shapes on the basis of
  individual exposures are described in detail in \citet{mhk12}. The
  quality of the shear estimates is discussed in \citet{hvm12}.
\item Each \CFHTLenS science image is supplemented by a mask,
  indicating regions within which accurate photometry/shape
  measurements of faint sources cannot be performed, e.g. due to
  extended haloes from bright stars.
\end{enumerate}
The methods and algorithms used to obtain the imaging products are heavily
based on our developments within the \CARS project
\citep[see][]{ehl09}. In the following we give a thorough description
of the steps that contain significant changes and improvements.
The main differences concern data treatment on the patch-level within
\CFHTLenS; while for \CARS we treated each survey pointing
independently we now simultaneously treat all images within a patch.
This optimally utilises available information to obtain a
homogeneous astrometric and photometric calibration over the patch area.
Our data processing is described in the following.
\subsection{Data Retrieval from \CADC}
\label{sec:dataretrieval}
\begin{figure}
  \centering
  \includegraphics[width=0.95\columnwidth]{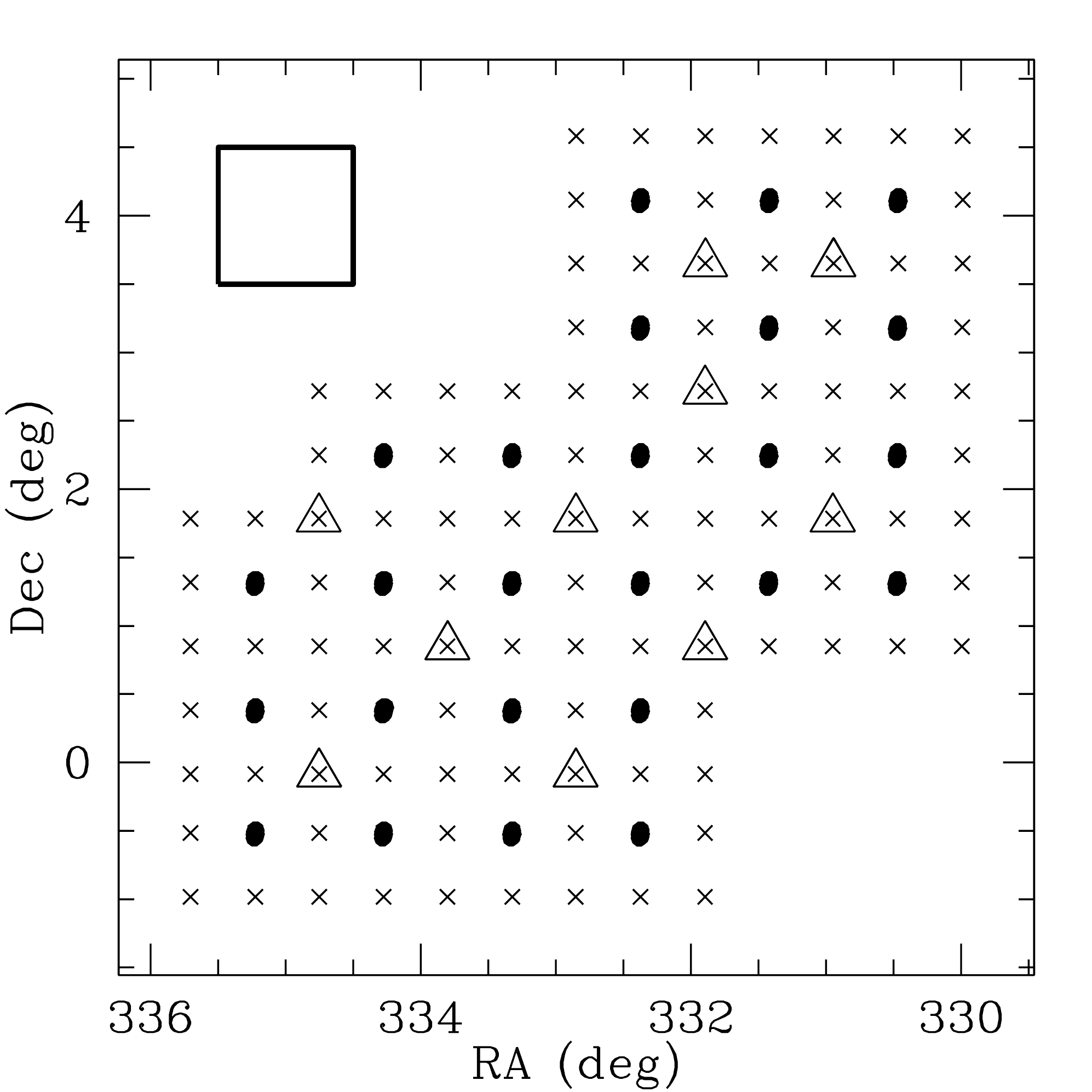}
  \caption{\label{fig:patchlayout} Available data in the \Wfour patch
    area: dots denote the centres of primary science observations,
    crosses indicate the centres of exposures of the astrometric presurvey and
    triangles mark the centres of additional photometric pegs.  The square
    in the upper left corner shows the \MegaPrime field-of-view.}
\end{figure}

We start our analysis with the
\Elixir\footnote{\url{http://www.cfht.hawaii.edu/Instruments/Elixir/}}
preprocessed \CFHTLS-Wide data available at the Canadian Astronomical
Data Centre
(\CADC)\footnote{\url{http://www4.cadc-ccda.hia-iha.nrc-cnrc.gc.ca/cadc/}}.
Exposure lists for the \CFHTLS surveys can be obtained from
\CFHT\footnote{
  \url{http://www.cfht.hawaii.edu/Science/CFHTLS-DATA/exposureslogs.html}}.
Besides the primary \CFHTLS-Wide imaging data the catalogue lists, for
each patch, exposures of an astrometric \emph{presurvey}. This
presurvey densely (re)covers the complete patch area with short (180s)
$r'$-band exposures. The footprint for the presurvey fields is
different from the science pointings to enable a good mapping of
camera distortions. At the end of the survey each patch was similarly
complemented with additional exposures obtained under photometric
conditions in all filters. Each of these photometric \emph{pegs}
overlaps with four science pointings and helps to ensure a homogeneous
photometric calibration on the patch
level. Figure~\ref{fig:patchlayout} outlines the available data for
patch \Wfour. The photometric pegs were not obtained under the primary
\CFHTLS programme but under the \CFHT programme IDs 08AL99 and
08BL99. Using the relevant exposure IDs all data were retrieved from
\CADC. Besides the image list, the \CFHTLS exposure catalogue also
contains information on the conditions of the observations.  Only data
that are marked as either \emph{completely within survey
  specifications} or as \emph{having one of the predefined
  specifications (seeing, sky transparency or moon phase) slightly out
  of bounds}\footnote{The conditions imposed on \CFHTLS-Wide
  observations were: image quality (seeing)$\leq 0\myarcsec 9$ for all
  filters, dark sky for $u^*$ and $g'$ observations and dark/gray moon
  phases for $r'$, $i'$ and $z'$ images. Thin cirrus was accepted for
  the complete science campaign (Cuillandre, private communication).}
enter the following process. We note that the availability of this
quality information made laborious quality checks on each image
unnecessary at this stage.
\subsection{Processing of single exposures}
\label{sec:singleexpprocessing}
In addition to raw data, \CADC offers all \CFHTLS images in \Elixir
preprocessed form.  The \Elixir processing \citep[see][]{mac04}
includes removal of instrumental signatures. This spans overscan-
and bias subtraction, flatfielding, removal of fringing in $i'$ and
$z'$, and photometric flattening across the \MegaPrime
field-of-view. In addition, each exposure comes with photometric
calibration information (zeropoint, extinction coefficient and colour
term)\footnote{See the \CFHT web pages
  \url{http://www.cfht.hawaii.edu/Science/CFHTLS-DATA/dataprocessing.html}
  and
  \url{http://www.cfht.hawaii.edu/Science/CFHTLS-DATA/megaprimecalibration.html}
  for a more detailed description of the \Elixir processing on \CFHTLS
  data.}.

Starting from the \Elixir images, we perform the following processing
steps \citep[see][for more details]{ehl09}: (1) we identify and mark
individual exposure chips that should not be considered any further
using a FITS header keyword. This concerns chips that either contain
no information (all pixel values equal zero) or are completely
dominated by saturated pixels from a bright star. In contrast to
\CARS we do not automatically mark chips in other colours of a
pointing as bad if the corresponding $i'$-band chip is flagged; (2) we
create sky-subtracted versions of all chips with \SExtractor; (3) we
create a weight image for each science chip as outlined in
\citet{esd05} and in detail for \MegaPrime data in Sect. A.2 of
\citet{ehl09}. As described in these publications, we aim for a
complete identification of image artefacts on the level of individual
chips to perform a weighted-mean co-addition of the data
later-on. Cosmic rays in our data are detected with a neural network
algorithm that utilises \SExtractor with a special cosmic ray filter.
This filter is constructed with the \EyE program\footnote{See
  \url{http://www.astromatic.net/software/eye}. \EyE produces
  detection filters for \SExtractor. It is a neural network classifier
  specialised to be trained for the detection of small scale features
  in imaging data. A filter for cosmic rays can be obtained by using
  image simulations or real data with cosmic rays imposed on known
  image positions. Cosmic ray like features themselfs can be extracted
  from long exposed dark frames for instance.  The \MegaPrime \EyE
  cosmic ray filter that we use for our analysis can be downloaded
  from \url{http://www.astromatic.net/download/eye/ret/megacam.ret}}
\citep[see][]{ber01b}.  In the course of our analysis we noted a
significant confusion of stellar sources with cosmic rays in images
obtained under superb seeing conditions. The effect is highly notable
for a seeing below $\sim 0\myarcsec 6$.
In \sectionref{sec:cosmicrays} we describe in detail how this
confusion is treated; (4) Utilising the weight image we extract
reliable, high $S/N$ object catalogues from each chip (\SExtractor
DETECTION\_MINAREA / DETECTION\_THRESH is set to 5 / 5 for
$g'r'i'y'z'$ and to 3 / 3 for $u^*$), which are used for our
astrometric and photometric calibration. (5) Finally, we study the PSF
properties of each chip by analysing bright, unsaturated stars with
the Kaiser-Squires and Broadhurst (KSB) algorithm
\citep[see][]{ksb95}.  This is done primarily to reject images with
badly behaved PSF properties such as a large stellar ellipticity at a
later stage, see \sectionref{sec:asphotcalib}.
\subsection{Astrometric and Photometric Calibration}
\label{sec:asphotcalib}
\begin{figure}
  \centering
  \includegraphics[width=0.95\columnwidth]{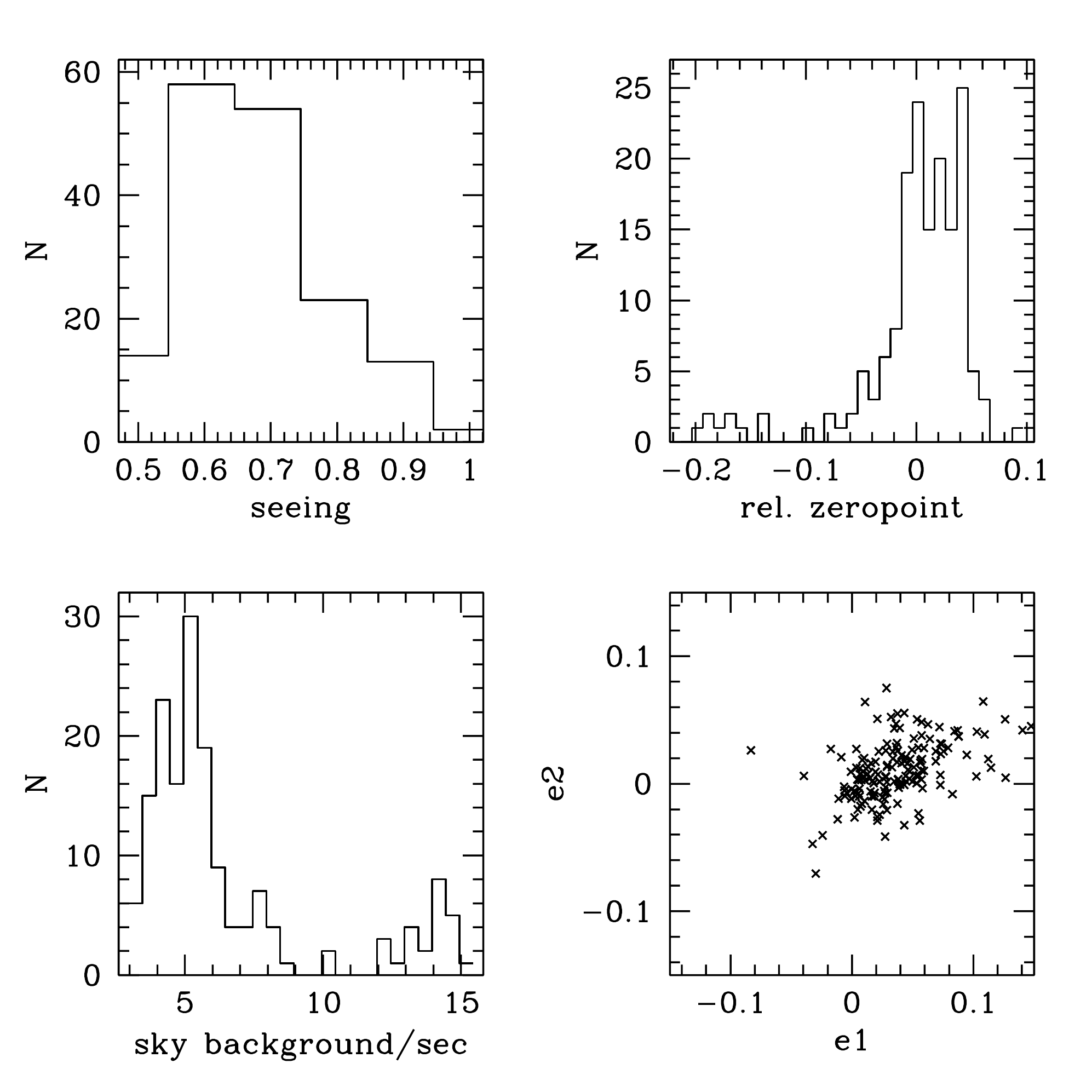}
  \caption{\label{fig:quality}
    Quality parameter distributions of all 164 \Wfour $i'$-band
    exposures that enter the co-addition and science analysis stage.
    Shown are the seeing distribution (top left), the distribution
    of relative photometric zeropoints as determined by \scamp
    (top right), the sky-background brightness in ADU/s (bottom left)
    and the two components of stellar PSF ellipticities (bottom right).
    All quantities are estimated as mean values over all 36 chips of
    a specific exposure. See the text for further details.}
\end{figure}

The most significant difference between the \CARS and the \CFHTLenS
data processing concerns the astrometric and photometric
calibration. While we treated each pointing separately and
independently in \CARS, we now perform these calibration steps
simultaneously for all exposures of a patch within \CFHTLenS. By
treating all available data at the same time we expect an increased
homogeneity in astrometric and photometric properties of the data. The
main pillar of this processing unit is the \scamp
programme\footnote{\url{http://www.astromatic.net/software/scamp}}
\citep[see][]{ber06}, which is
specifically designed for accurate astrometric and photometric
calibration of large imaging surveys. The size of the survey that
can be calibrated with \scamp in a single step is only limited by
computational resources, especially the main memory. We perform the
following calibration steps:
\begin{enumerate}
\item Our astrometric reference catalogues are \TWOMASS
  \citep[see][]{scs06} for \Wone, \Wtwo and \Wfour and \SDSSRS
  \citep[see][]{aaa09} for \Wthree. Unfortunately, the \SDSSRS only
  covered patch \Wthree completely and small parts of the other
  \CFHTLenS areas.
\item The available computer equipment\footnote{Our main processing
    machine is a 48 core AMD Opteron Processor (with a clock rate of
    2100 MHz) computer installed at the University of British
    Columbia. The machine is equipped with 128GB of main memory from
    which we separate 100GB for a RAM disk. The RAM disk allows us to
    perform time-dominant I/O operations within the physical memory
    and to reach a high machine work load for nearly the complete
    data processing cycle.}  allowed us to calibrate all exposures
  (primary science, astrometric presurvey, photometric pegs) from all
  filters of the smaller patches \Wtwo and \Wfour simultaneously.
  Both patches consist of about 1000 individual \MegaPrime
  exposures with 36 chips each. The larger patches \Wone ($\sim$3000
  exposures) and \Wthree ($\sim$2000 exposures) had to be split for
  our \scamp runs. First, we separately process the $r'$-filter, which
  consists of science data in addition to the astrometric presurvey
  images. Next, the remaining filters $u^*$, $g'$, $i'$ and $z'$ were
  individually calibrated together with the $r'$-band, so that each
  filter profited from the astrometric presurvey information. In
  addition to astrometric calibration, \scamp uses sources from
  overlapping exposures to perform a \emph{relative} photometric
  calibration. For each exposure, $i$, of a specific filter, $f$, we
  obtain a relative magnitude zeropoint, $ZP_{\rm rel}(i, f)$,
  giving us the magnitude offset of that image with respect to the mean
  relative zeropoint of all images. That is, we demand $\sum_i ZP_{\rm
    rel}(i, f) = 0$.  Note that this procedure calibrates data
  obtained under photometric and non-photometric conditions on a
  relative scale.  An absolute flux scaling for the patch can be
  obtained from the photometric subset; see below\footnote{ \scamp
    offers the possibility to internally perform a complete absolute
    photometric calibration and to finally calibrate/rescale all data
    to a predefined absolute magnitude zeropoint. The \scamp default
    for this zeropoint is 30. We do not make use of this feature,
    mainly to be consistent with the original \THELI data-flow
    \citep[see][]{esd05,ehl09} and to preserve a \emph{standard}
    scaling (ADU/s) for the pixel values of our co-added images.}.
\item After the first \scamp run we reject exposures suffering from an
  atmospheric extinction larger than 0.2 mag.  We also remove images
  showing a large PSF ellipticity over the field-of-view. Large,
  homogeneous PSF anisotropies are mostly a sign of tracking problems
  during the exposure. All images that have a mean stellar ellipticity
  (the mean is taken over all chips of the image and it is estimated
  with the KSB algorithm) of 0.15 or larger are
  discarded from further analyses. Utilising the remaining images, we
  perform another \scamp run to conclude the astrometric and relative
  photometric calibration of our data. For each patch and filter
  we manually verify the distributions of typical quality
  parameters (sky-background level, seeing, stellar ellipticity,
  relative photometric zeropoint). None of the plots showed suspicious
  images that should be removed at this stage. See
  \figref{fig:quality} for an example of our patch-wide check plots.
\item The last step of the astrometric and photometric calibration is
  the determination of the absolute photometric zeropoint on the patch
  level.  Input to our procedure are the relative zeropoints from
  \scamp, photometric zeropoints and extinction coefficients from
  \Elixir, and the list of exposures that were obtained under
  photometric conditions. Information on the sky-transparency of each
  image is included in the \CFHTLS exposure catalogue
  (see \sectionref{sec:dataretrieval}). For all photometric exposures,
  $i$, in a filter, $f$, from a given patch, we calculate a
  \emph{corrected zeropoint}, $ZP_{{\rm corr}}(i, f)$, according to
  \[
  ZP_{{\rm corr}}(i, f) = ZP(i, f) + AM(i, f) EXT(i, f) + ZP_{{\rm
      rel}}(i, f),
  \]
  where $ZP(i, f)$ is the instrumental AB zeropoint, $AM(i, f)$ the
  airmass during observation, and $EXT(i, f)$ is the colour-dependent
  extinction coefficient. For photometric data, the relative
  zeropoints compensate for atmospheric extinction and the corrected
  zeropoints agree within measurement errors. We iteratively estimate
  the mean $ZP(f) = \langle ZP_{{\rm corr}}(f)\rangle_i$ of all
  exposures, $i$, by rejecting 3-$\sigma$ outliers.  With more than
  100 exposures marked as photometric in each patch and filter, this
  procedure ensures a robust estimation of the patch zeropoint.  The
  final $ZP(f)$ is used as the absolute magnitude zeropoint for all
  co-added images of filter, $f$, in a particular patch.
\end{enumerate}
We assess the quality of our astrometric and photometric calibration in
\sectionref{sec:astromphotomqual}.
\subsection{Image Co-Addition and Mask Creation}
\label{sec:coaddmask}
In the subsequent analysis, co-added data are used in the detection of
stars and galaxies and in the photometric measurements and analysis
\citep[][]{hek12}.  Coadded data are not used for the lensing shear
measurement \citep[][]{mhk12}.  One of our main goals for the coadded
images is to ensure data with homogeneous image quality. We therefore
check for each pointing/filter combination whether the exposure set
consists of images with large seeing variations. For instance our best
seeing pointing W4m3p1 $i'$-band has a co-added image seeing of
$0\myarcsec 44$ though originally it has four individual exposures
with image qualities of $0\myarcsec 43$, $0\myarcsec 47$, $0\myarcsec
48$ and $0\myarcsec 88$.  To avoid \emph{degradation} of the superb
quality images below $0\myarcsec 5$ with the image of $0\myarcsec 88$
we want to reject the last image from the co-addition process. We
estimate the median ($med$) of the seeing values of a pointing/filter
combination and reject data that have a larger seeing than $med +
0.25$. In addition, for the $i'$-band data, which form the basis for
our source catalogues, images with a seeing larger than $1\myarcsec 0$
are not included in the co-addition process.  Note that our procedure
ensures homogeneity on the pointing/filter level and avoids
rejection of data with fixed quality values on the patch
level\footnote{It is important to stress that the seeing selection for
  our co-added images is not propagated to the \lensfit shear
  analysis, which is based on joint analysis of individual exposures
  \citep{mhk12}. All $i'$-band exposures that have not been rejected by the
  end of the astrometric and photometric calibration process enter the
  \lensfit shear analysis.}.

Finally, the sky-subtracted exposures belonging to a pointing/filter
combination are co-added with the \swarp
programme\footnote{\url{http://www.astromatic.net/software/swarp}}
\citep[see][]{ber02b}.  We use the LANCZOS3 kernel to remap original
image pixels according to our astrometric solutions.  The subsequent
co-addition is done with a statistically optimally weighted mean which
takes into account sky-background noise, weight maps and the relative
photometric zeropoints as described in Sect. 7 of \cite{esd05}. As sky
projection we use the TAN projection \citep[see][]{grc02}.  The
reference points of the TAN projection for each pointing are those
defined for the \CFHTLS-Wide survey\footnote{see
  \url{http://terapix.iap.fr/cplt/oldSite/Descart/summarycfhtlswide.html}}.
After co-addition we cut all images to a common size of $21k\times
21k$ around the image centre. This cut comprises areas with useful data
for all \CFHTLenS pointings.
The \swarp information and photometric zeropoints are also passed to
the lensing shear analysis of the individual exposures, although a key
part of the shear measurement is that the data are not interpolated
onto a new reference frame when measuring galaxy shapes \citep[][]{mhk12}.

As a final step we use the \automask tool\footnote{
  \url{http://marvinweb.astro.uni-bonn.de/data_products/THELIWWW/automask.html}}
\citep[see][]{del07} to create image masks for all pointings. These
masking procedures are described in detail in \citet{ehl09}.  Within
\CFHTLenS all 171 automatically generated masks are manually
double-checked and, if necessary, refined. We note that the lensing
catalogue quality assessment performed in \citet{hvm12} shows that
lensing analyses with the automatic masks and the refined versions are
consistent.

The result of this step are co-added science images for all 171
\CFHTLenS pointings in all filters. Each science image is accompanied by
a \emph{weight} and a \emph{sum} image as described in
\sectionref{sec:dataprocessing}.  These products, together with the
sky-subtracted individual chip data and the astrometric information
from \scamp (see \sectionref{sec:asphotcalib}) form the basis for all
\CFHTLenS shear and photometric analyses.
\section{Influence of our Cosmic Ray removal on stellar sources}
\label{sec:cosmicrays}
\begin{figure}
  \centering
  \includegraphics[width=0.95\columnwidth]{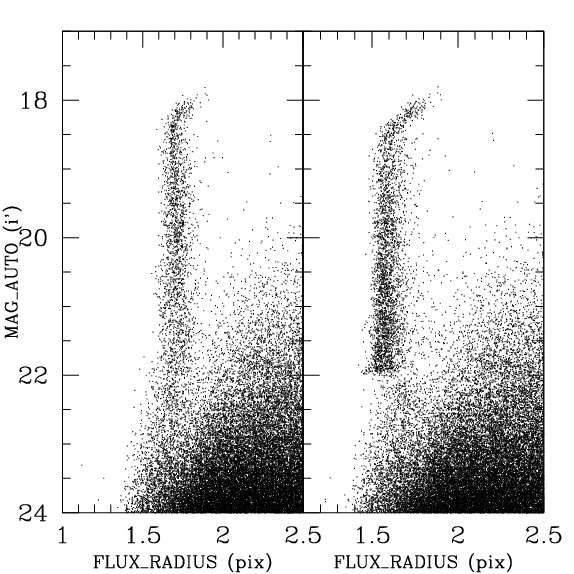}
  \includegraphics[width=0.95\columnwidth]{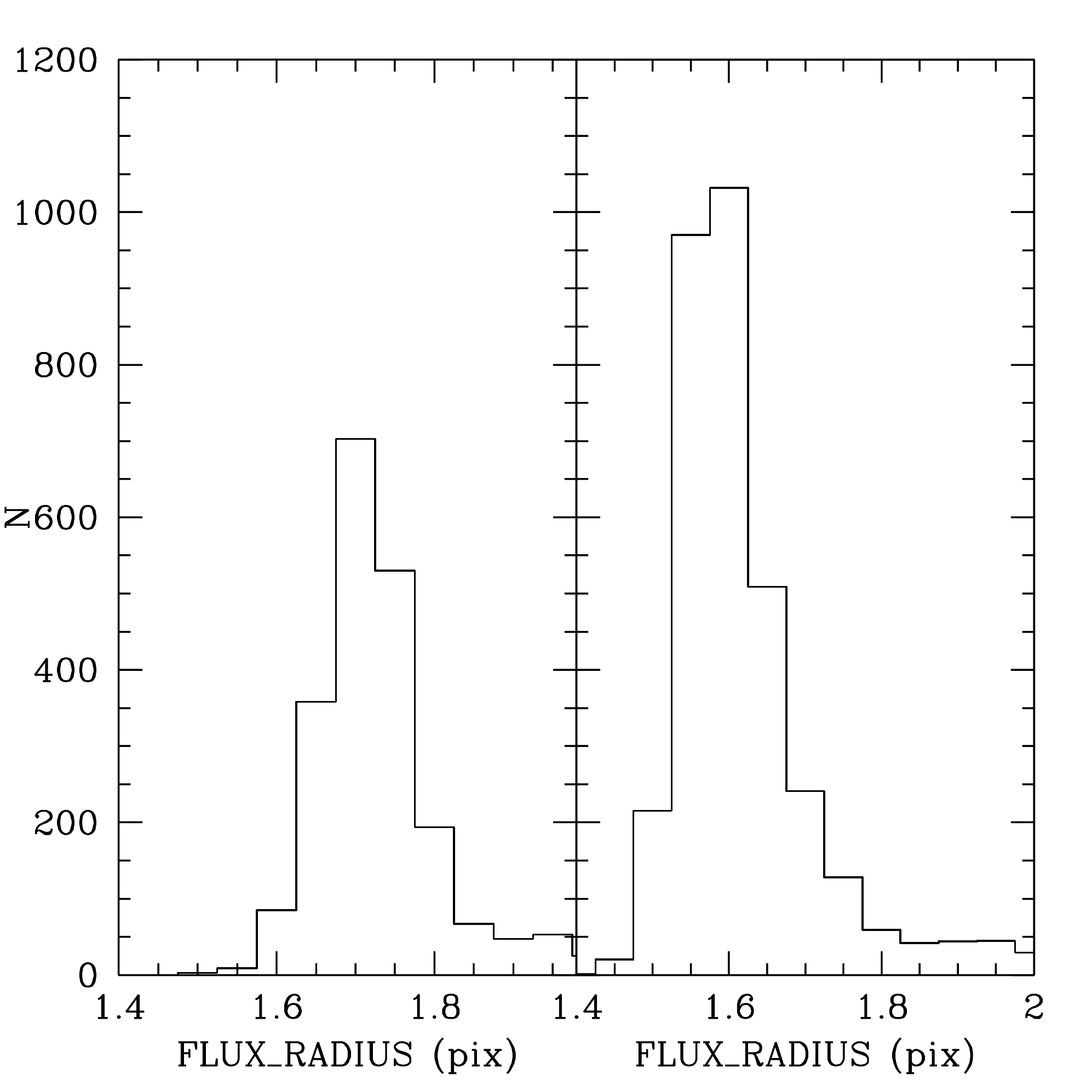}
  \caption{\label{fig:stellar_break} \emph{Stellar break} in the
    co-added image of W1m2m1 $i'$-band, with a seeing of $0\myarcsec
    47$: Shown are stellar loci in the size-mag plane (\SExtractor
    quantities FLUX\_RADIUS and MAG\_AUTO; top panels). The top
    left panel shows the stellar locus after our standard cosmic-ray
    removal procedure, the top right panel after we bring back stars
    whose cores were falsely classified as cosmic rays. The lower
    panels show corresponding histograms of object counts for
    $1.4<\mbox{\texttt{FLUX\_RADIUS}}<2.0$ and $i'<22.0$. See text for
    further details.}
\end{figure}

\begin{figure}
  \centering
  \includegraphics[width=0.95\columnwidth]{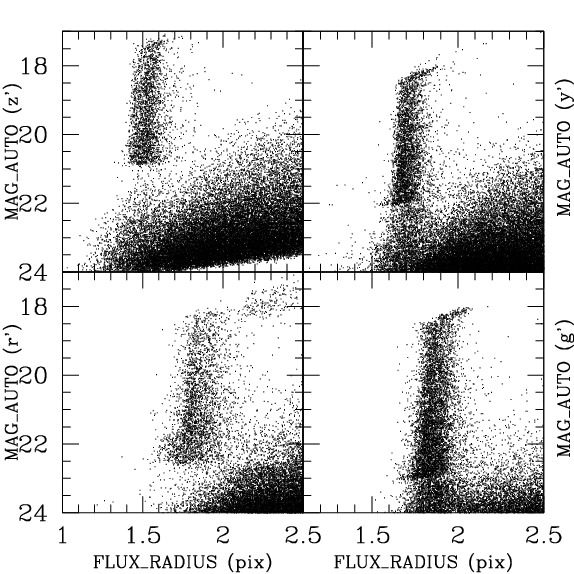}
  \caption{\label{fig:sb_colours}
    \emph{Stellar break} in {W1p4p1} $z'$-band ($0\myarcsec 46$,
    top left),
    {W3m2m1} $y'$-band ($0\myarcsec 51$, top right),
    {W1p4p1} $r'$-band ($0\myarcsec 52$, bottom left)
    and {W4p1p1} $g'$-band ($0\myarcsec 58$, bottom right);
    see text for further details.}
\end{figure}

As discussed in \sectionref{sec:singleexpprocessing}, our procedure to
identify cosmic rays in individual \MegaPrime exposures is based on a
neural network approach. During the weak lensing analysis with
\lensfit we noticed that a large number of individual exposures had
very few stars suitable for PSF analysis. We traced the problem to the
cores of point sources being misclassified and masked as cosmic
rays. A closer analysis revealed that the problem was worst for the
best seeing exposures and the neural network approach the primary
source of the problem. In the following our main goal is to unflag
bright, unsaturated stars suitable for PSF analyses with \lensfit and
PSF homogenisation within our photometric redshift (photo-$z$)
analyses \citep[see][]{hek12}. We explicitly note that we did not aim
for a complete solution to the problem within \CFHTLenS. Our
prescription to identify and to unflag bright stars after the initial
cosmic ray analysis is as follows: (1) We run \SExtractor on
individual exposure chips with a high detection threshold
(DETECTION\_MINAREA / DETECTION\_THRESH is set to 10 / 10). This
\SExtractor run is performed without using weighting or flagging
information. (2) Candidate stellar sources are identified on the
stellar locus in the size-magnitude plane. (3) We perform a standard
PSF analysis with the KSB algorithm. This involves estimating weighted
second-order brightness moments for all candidate stars and to
perform, on the chip level, a two-dimensional second order polynomial
fit to the PSF anisotropy.  The fit is done iteratively with outliers
removed to obtain a clean sample of bright, unsaturated stars suitable
for PSF analysis.  (4) We remove cosmic ray masks in a square of $4x4$
pixels around stellar sources that are still included in our sample
after step (3).  Figure~\ref{fig:stellar_break} shows the result of
our analysis on pointing W1m2m1 in the $i'$-band.  The set consists of
seven exposures with an image quality between $0\myarcsec 48$ and
$0\myarcsec 55$, including five images below $0\myarcsec 5$.
The figure shows the stellar locus of the co-added image before (left
panel) and after (right panel) we modified the cosmic ray masks of
individual exposures.  We note that our procedure returns a
significant number of stars to the sample. In the corrected version we
also see an abrupt break in the stellar locus at $i'\approx 22$. For
our $i'$-band data this marks the limit to identify usable stars for
PSF studies with our KSB approach, and we would need another procedure
to also reliably  identify fainter stars that are confused as cosmic
rays. We would like to reiterate that our main goal within \CFHTLenS
is to have a sufficient number of bright, unsaturated stars for a
reliable PSF analysis with \lensfit, but none of our science projects
requires complete and unbiased stellar samples down to faint
magnitudes.  We identified the \emph{stellar break} problem to be
immediately noticeable in images with a seeing of about $0\myarcsec 6$
and better.  This feature is more prominent the better the image
quality is. In the co-added images with overall seeing of
$0\myarcsec 7$ - $0\myarcsec 75$ we can still identify stellar breaks
if the set contains exposures in the best seeing range. In
\figref{fig:sb_colours} we show prominent stellar breaks for
$i'\approx 22$, $z'\approx 21$, $r'\approx 22.5$ and $g'\approx 23$.
We do not observe obvious breaks in the loci of $u^*$, where the best
quality coadd has an image seeing of $0\myarcsec 62$, and only some in
$g'$.  Fields with obvious stellar breaks are indicated in the
comments column of \tabref{tab:CFHTLenSquality}. The judgement was
done subjectively by manually checking stellar locus plots from all
171 \CFHTLenS pointings.  We specifically note that our cosmic ray
removal procedure did not influence the detection nor the photometry
of galaxies.
\section{Evaluation of Astrometric and Photometric Properties}
\label{sec:astromphotomqual}
Our data underwent substantial testing and quality control for our
main scientific objective: weak gravitational lensing studies with
photometric redshifts for all galaxies.  The
quality of our \lensfit shear estimates and the accuracy of
photometric redshifts are described in detail in \citet{hvm12} and
\citet{hek12}. These analyses have demonstrated the robustness of our
data set. Here we mainly quote the precision we were able to achieve
in our astrometric and photometric calibration.

\begin{figure}
  \centering
  \includegraphics[width=0.95\columnwidth]{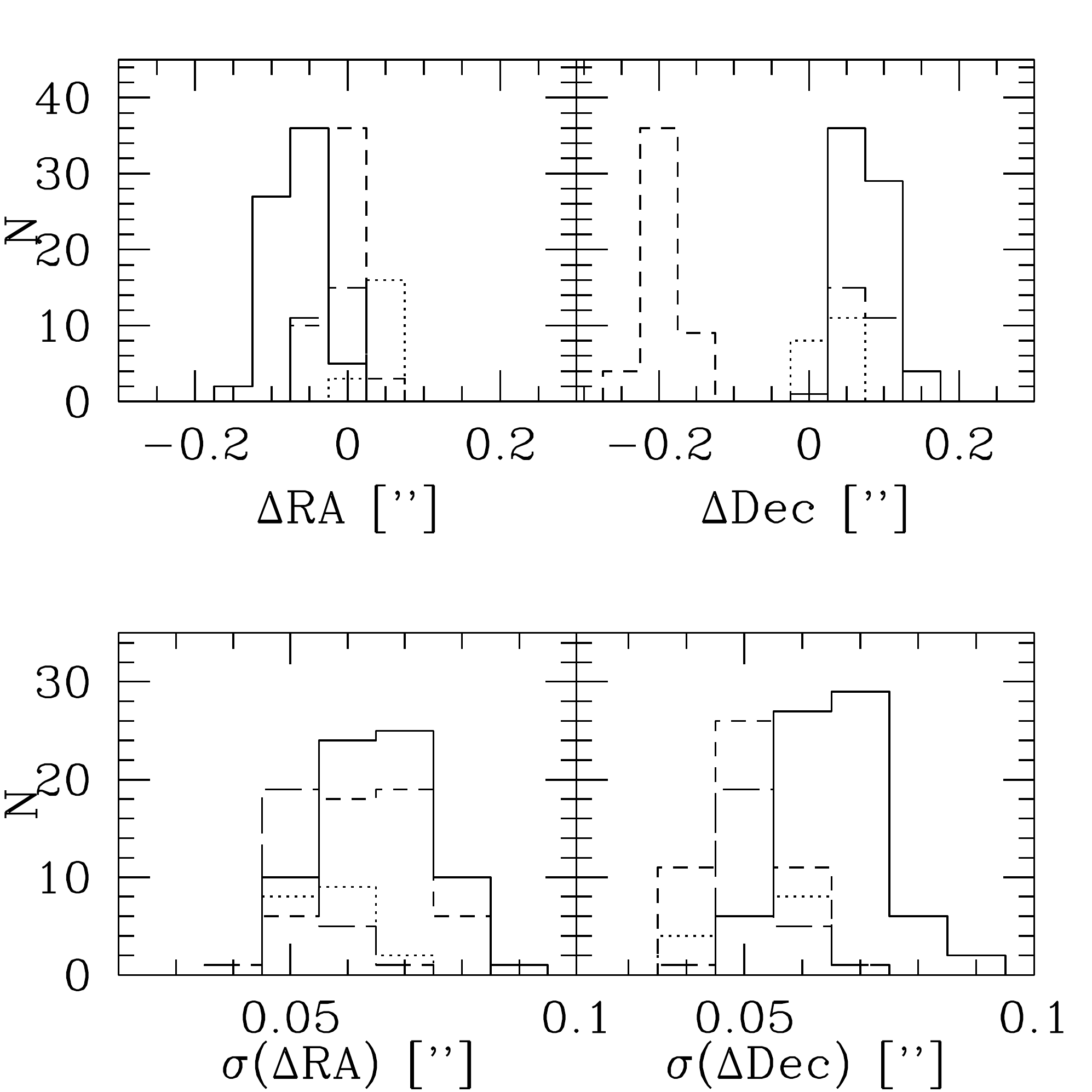}
  \caption{\label{fig:sloan_external} Astrometric comparison with \SDSSRE:
  Shown are object position comparisons between \CFHTLenS sources in all
  pointings and all filters with \Sloan $i_{\rm Sloan} < 21$ stars.
  Solid, dotted, short-dashed and long-dashed histograms show
  comparisons of \Wone, \Wtwo, \Wthree and \Wfour respectively. See
  text for further details.}
\end{figure}

To quantify our astrometric accuracy with respect to external sources
we compare object positions in our \CFHTLenS pointings with the
\SDSSRE catalogue \citep[see][]{aaa11}. Note that \SDSSRE\footnote{
  \SDSSRE is a complete reprocessing of the entire \Sloan data with
  improved processing techniques (\url{http://www.sdss3.org/dr8/}).
  It is therefore also an independent test set for \Wthree which was
  astrometrically calibrated with \SDSSRS.} was not used as an
external astrometric catalogue for our astrometric calibration. It
only became available after our data processing was completed. It is
the first \Sloan catalogue that covers all but ten \CFHTLenS
pointings. The fields without \SDSSRE overlap are {W1p3m4}, {W1p4m4}
and the eight {W2} pointings south of $-4$ degrees in declination (see
\figref{fig:cfhtlenslayout}). Figure~\ref{fig:sloan_external}
summarises our astrometric accuracy compared to the \Sloan
reference. We compare the position of \Sloan stellar sources with
$i_{\rm \Sloan} < 21$ to each pointing and filter . Object positions
in our data were estimated independently for each filter in the
corresponding co-added images.  The star classification was taken
from the \Sloan catalogue. Figure~\ref{fig:sloan_external} shows the
mean deviation (the mean is taken over all sources in all filters in a
patch) of positions and the standard deviation of the positional
differences. We see that the \CFHTLenS data show a systematic offset
in right ascension and declination of less than $0\myarcsec 2$ in all
cases but one. We note however that the \Sloan team acknowledges a
systematic offset of 250 mas in declination for $\mbox{Dec} > +41$
degrees in the \SDSSRE catalogue\footnote{
  \url{http://www.sdss3.org/dr8/algorithms/astrometry.php#caveats}}. This
affects patch \Wthree at a declination of $\mbox{Dec}\approx +54$ deg.
The standard deviation is uniform over all fields and the
its distribution peaks at about 50-70 mas for all \CFHTLenS patches.

\begin{figure}
  \centering
  \includegraphics[width=0.95\columnwidth]{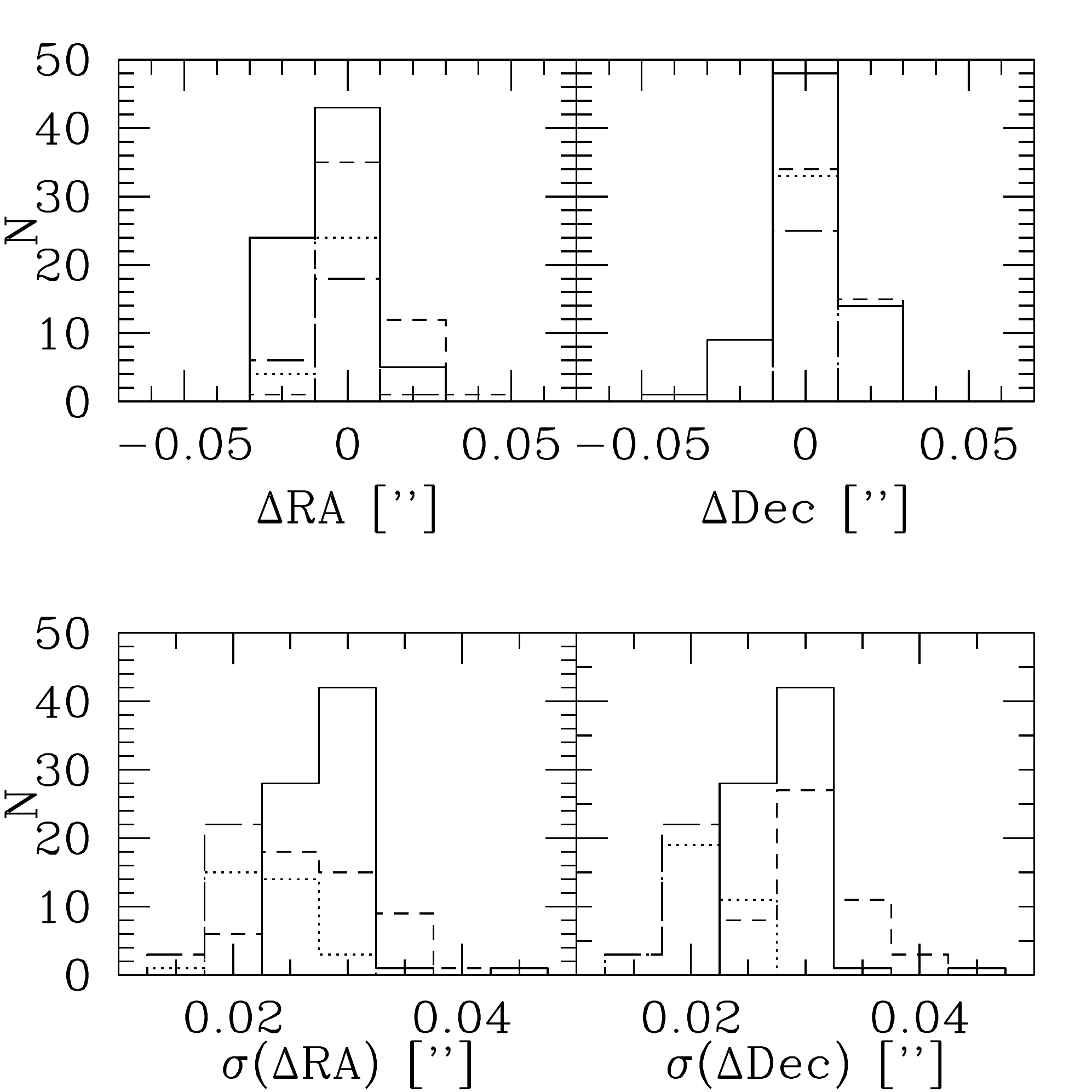}
  \caption{\label{fig:internal_astrom} Internal astrometric accuracy:
  Shown are internal astrometric positional differences between the
  different filters \emph{within} individual \CFHTLenS pointings.
  Solid, dotted, short-dashed and long-dashed histograms show
  comparisons of \Wone, \Wtwo, \Wthree and \Wfour respectively. See
  text for further details.}
\end{figure}

\begin{figure}
  \centering
  \includegraphics[width=0.95\columnwidth]{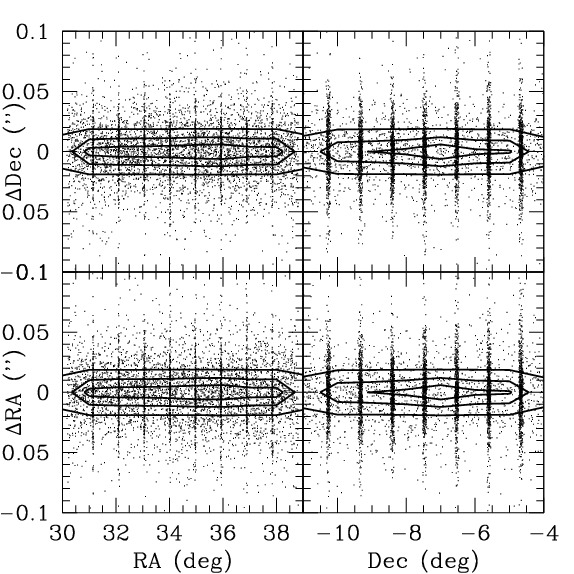}
  \caption{\label{fig:internal_astrom_overlap} Internal astrometric
    accuracy on overlap sources in \Wone: We show positional
    differences between object matches of \CFHTLenS sources in
    \emph{different}, adjacent pointings.  The comparison is done in
    \Wone across all filters. Vertical stripes in the density
    distribution originate from the alignment of overlap
    regions; see \figref{fig:cfhtlenslayout}. Contours indicate areas
    of 0.7, 0.4, and 0.05 times the peak-value of the point-density
    distribution. For clarity of the plot, only 1 point out of
    100 is visualised. See text for further details.}
\end{figure}

In Figs.~\ref{fig:internal_astrom} and
\ref{fig:internal_astrom_overlap} we quantify the internal astrometric
accuracy, comparing positions of sources observed in different
filters of all pointings. We use objects with $i'_{\CFHTLenS} < 21$
that are classified as stars by \SExtractor
(\texttt{CLASS\_STAR}$>0.95$). The sources were extracted from the
co-added images. Figure~\ref{fig:internal_astrom} shows positional
differences \emph{within} individual \CFHTLenS pointings. We see that
we cannot detect significant systematic offsets in right ascension and
declination between the colours.  The rms positional difference
between the filters is about 30 mas.  In
\figref{fig:internal_astrom_overlap} we show positional differences
with sources on \emph{different} \CFHTLenS pointings. As before, we
match objects regardless of their filter, but only allow
associations from different, adjacent \CFHTLenS pointings. We only
show the {W1} comparison here -- results are similar for the
other patches.  The error parameters are comparable to the
\emph{inter-pointing} comparison. Absolute positional differences are
evenly distributed around zero and the rms deviations are
$\sigma(\Delta {\rm RA}) = 0\myarcsec 030$ and $\sigma(\Delta {\rm
  Dec}) = 0\myarcsec 027$.

The photometric calibration of \CFHTLenS is also evaluated by direct
comparison to \SDSSRE.  The availability of \Sloan data nearly
overlapping the full \CFHTLenS area allows us to obtain a
comprehensive understanding of the photometric quality of our data. We
would like to reiterate that the \Sloan data were not used at any
stage of the data calibration phase.

\begin{figure}
  \centering
  \includegraphics[width=0.95\columnwidth]{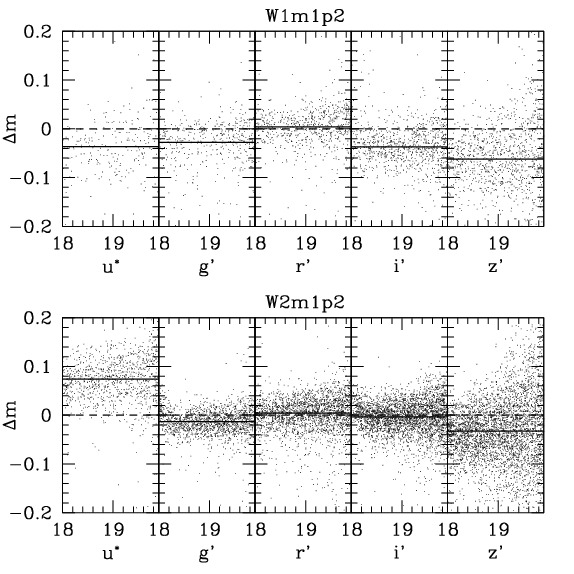}
  \includegraphics[width=0.95\columnwidth]{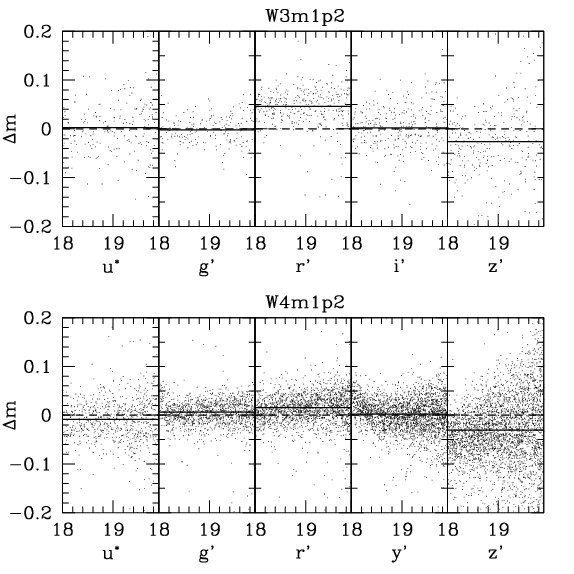}
  \caption{\label{fig:sloan_comp_m1p2} Magnitude comparisons between
    \Sloan stars with and \CFHTLenS sources for the fields
    \texttt{W$x$m1p2} with $x\in \{1,2,3,4\}$. Solid horizontal lines
    indicate $\langle \Delta \mbox{m}\rangle$.  The precise values of
    the mean offsets and formal standard deviations can be found in
    \tabref{tab:CFHTLenSquality}.  Note that \Wfour and \Wtwo are at
    significantly lower galactic latitude than \Wone and
    \Wthree, thus the stellar density in the latter two is
    substantially lower.}

\end{figure}

\begin{figure}
  \centering
  \includegraphics[width=0.95\columnwidth]{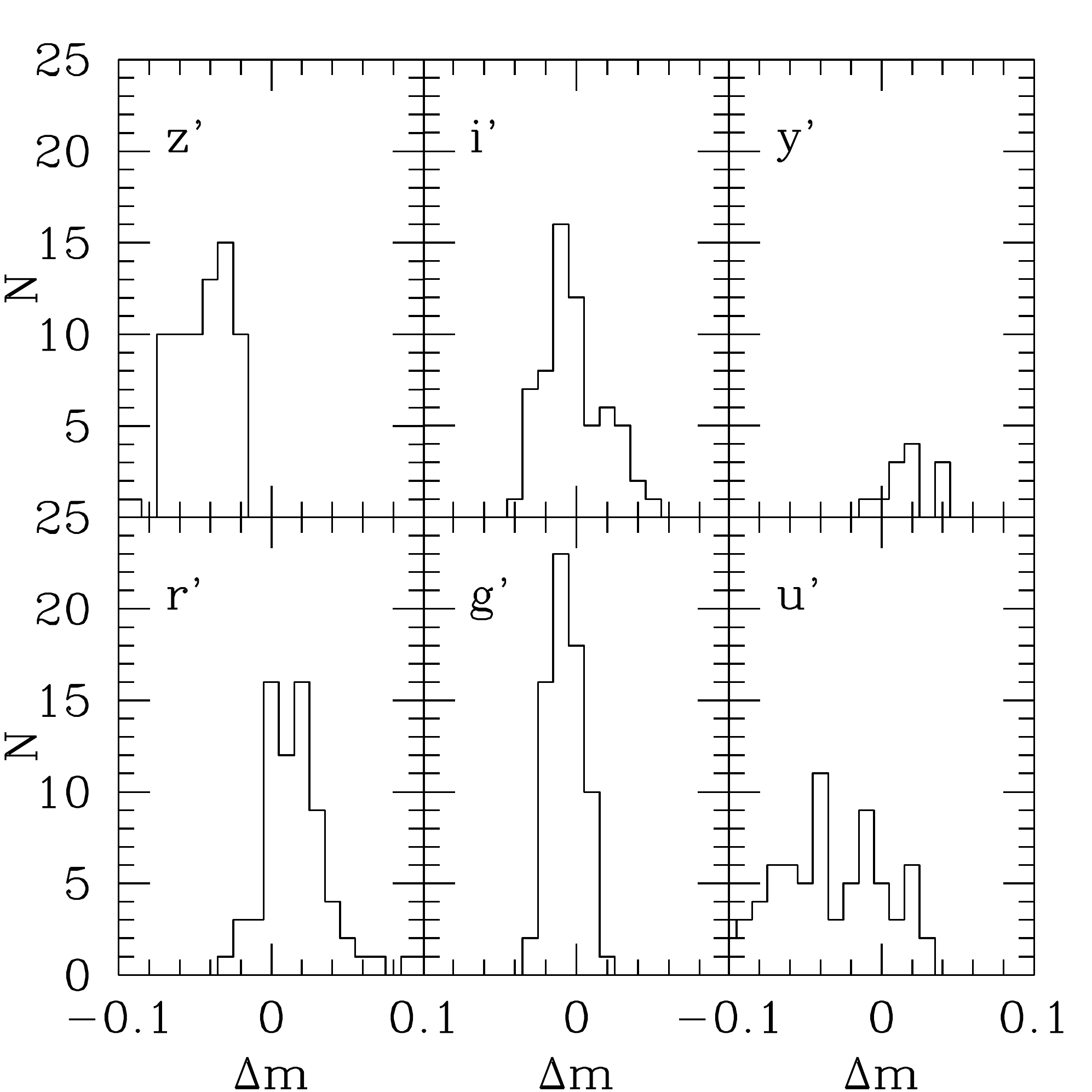}
  \caption{\label{fig:sloan_phot_mean} Distribution of the differences between
  \Sloan and \CFHTLenS magnitudes in \Wone. The abscissa of the plots show
  $\Delta m= m_{\rm \CFHTLenS}-m_{\rm \Sloan}$. See text for further details.}
\end{figure}

We compare \Sloan magnitudes of stellar objects with $i_{\rm \Sloan} < 21$
with their \CFHTLenS counterparts. To convert stellar \CFHTLenS AB
magnitudes to the \Sloan system we use the relations:
\begin{eqnarray}
\label{eq:sloantransform}
u^*_{\rm AB} & = & u_{\rm \Sloan} - 0.241 \cdot (u_{\rm \Sloan} - g_{\rm \Sloan}),
\nonumber \\
g'_{\rm AB}  & = & g_{\rm \Sloan} - 0.153 \cdot (g_{\rm \Sloan} - r_{\rm \Sloan}),
\nonumber \\
r'_{\rm AB}  & = & r_{\rm \Sloan} - 0.024 \cdot (g_{\rm \Sloan} - r_{\rm \Sloan}), \\
i'_{\rm AB}  & = & i_{\rm \Sloan} - 0.085 \cdot (r_{\rm \Sloan} - i_{\rm \Sloan}),
\nonumber \\
y'_{\rm AB}  & = & i_{\rm \Sloan} + 0.003 \cdot (r_{\rm \Sloan} - i_{\rm \Sloan}),
\nonumber \\
z'_{\rm AB}  & = & z_{\rm \Sloan} + 0.074 \cdot (i_{\rm \Sloan} - z_{\rm \Sloan}). 
\nonumber
\end{eqnarray}
The relations for $g'r'i'z'$ were determined within the \CFHTLS-Deep
Supernova project\footnote{see
  \url{http://www.astro.uvic.ca/~pritchet/SN/Calib/ColourTerms-2006Jun19/index.html#Sec04}};
the $u^*$ transformation comes from the CFHT instrument
page\footnote{see
  \url{http://cfht.hawaii.edu/Instruments/Imaging/MegaPrime/generalinformation.html}}
and the $y'$ equation was determined within the \MegaPipe
project\footnote{see
  \url{http://www3.cadc-ccda.hia-iha.nrc-cnrc.gc.ca/megapipe/docs/filters.html}}
\citep[][]{gwy08}. Magnitude comparisons on an object by object
basis for one randomly chosen field in each patch are shown in
\figref{fig:sloan_comp_m1p2}.  We see that the comparisons show a
dispersion of about $0.03-0.06$
magnitudes. Figure~\ref{fig:sloan_phot_mean} shows the distribution of
mean offsets in all pointings of the \Wone area. The results
are similar for the other patches.  The
offset distribution strongly peaks below $|\Delta m| \approx 0.04$ for
$g'$, $r'$, $i'$ and $y'$. It is significantly broader in $u^*$, and $z'$
peaks at around $\Delta m\approx -0.05$. As can be seen in
\figref{fig:sloan_comp_m1p2}, the relation between $z'_{\rm AB}$
and $z_{\rm \Sloan}$ leads to a significant spread on an object by object
basis. In rare cases we observe larger deviations between \Sloan and
\CFHTLenS magnitudes of up to $|\Delta m| \approx 0.1$. A detailed
list of the offsets for all \CFHTLenS fields with \Sloan overlap is
given in \tabref{tab:CFHTLenSquality}.

Given the results from the \SDSSRE comparison, we summarise
accuracies for the individual patches and filters in
\tabref{tab:photacc}. We quote the mean of all average deviations in
the individual pointings and their corresponding standard deviation.
The values indicate that we obtain on average a homogeneous
calibration of our data. This result is confirmed by the quality of our
photometric redshifts presented in \citet{hek12}. Since then we
were able to further test our photo-$z$ estimates with new
spectroscopic redshifts on a significant part of the \CFHTLenS
area. This additional confirmation for the robustness of our
photometry is described in the next section.
\begin{table}
  \caption{Average photometric accuracies in the \CFHTLenS patches}
\label{tab:photacc}
\centering          
\begin{tabular}{ccc|ccc}    
\hline\hline       
Patch & Filter & phot. accuracy & Patch & Filter & phot. accuracy \\ 
\hline
W1 & $u^*$ & $-0.034\pm 0.035$ & W1 & $i'$ & $-0.002\pm 0.020$ \\
W2 & $u^*$ & $+0.034\pm 0.031$ & W2 & $i'$ & $-0.009\pm 0.020$ \\
W3 & $u^*$ & $-0.046\pm 0.043$ & W3 & $i'$ & $+0.003\pm 0.015$ \\
W4 & $u^*$ & $-0.001\pm 0.014$ & W4 & $i'$ & $-0.003\pm 0.021$ \\
\hline
W1 & $g'$ & $-0.007\pm 0.011$ & W1 & $y'$ & $-0.019\pm 0.015$ \\
W2 & $g'$ & $+0.005\pm 0.013$ & W2 & $y'$ & $-0.021\pm 0.023$ \\
W3 & $g'$ & $-0.007\pm 0.012$ & W3 & $y'$ & $-0.002\pm 0.023$ \\
W4 & $g'$ & $-0.002\pm 0.010$ & W4 & $y'$ & $+0.022\pm 0.048$ \\
\hline
W1 & $r'$ & $+0.017\pm 0.024$ & W1 & $z'$ & $-0.045\pm 0.018$ \\
W2 & $r'$ & $+0.013\pm 0.012$ & W2 & $z'$ & $-0.054\pm 0.012$ \\
W3 & $r'$ & $+0.022\pm 0.014$ & W3 & $z'$ & $-0.040\pm 0.017$ \\
W4 & $r'$ & $+0.014\pm 0.006$ & W4 & $z'$ & $-0.030\pm 0.017$ \\
\hline
\end{tabular}
\end{table}
\subsection{Comparison of \CFHTLenS photo-$z$ with spectroscopic redshifts}
\label{sec:specz_comp}
The derivation of the \CFHTLenS photo-$z$ is detailed in \cite{hek12},
where we compared the photo-$z$ to spectroscopic redshifts (spec-$z$)
from \VVDS \citep{fvg05}, \DEEPT \citep{dgk07}, and \SDSSRS on 20 of
the 171 \CFHTLenS fields. More spec-$z$ have since become available
through the \VIMOS Public Extragalactic Redshift Survey (\VIPERS; see
Guzzo et al. 2013, in
preparation)\footnote{\url{http://vipers.inaf.it}}.  In this paper we
study how the \CFHTLenS photo-$z$ compare to \VIPERS on 22 additional
fields independent from the 20 fields tested in \cite{hek12}.

\begin{figure}
\includegraphics[width=0.95\columnwidth]{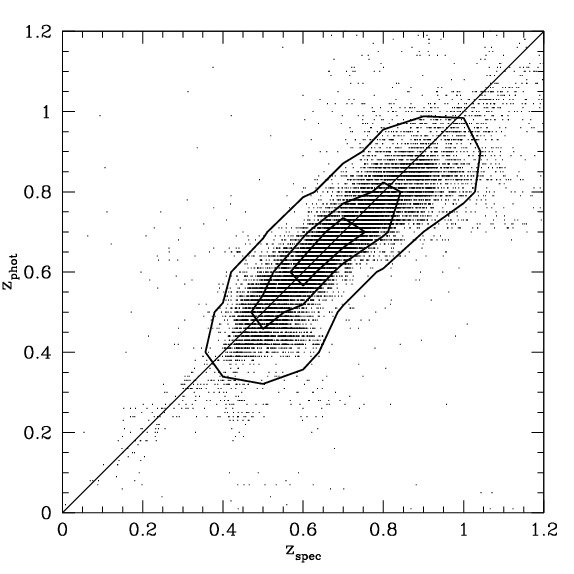}
\caption{Photo-$z$ vs. spec-$z$ for the 22 \CFHTLenS fields with \VIPERS
  overlap. Shown are all objects with secure spec-$z$. No magnitude
  cut is applied. Contours indicate regions around 0.7, 0.4, and 0.05
  times the peak-value of the point-density distribution.}
\label{fig:zz}
\end{figure}

\begin{figure}
\includegraphics[width=0.95\columnwidth]{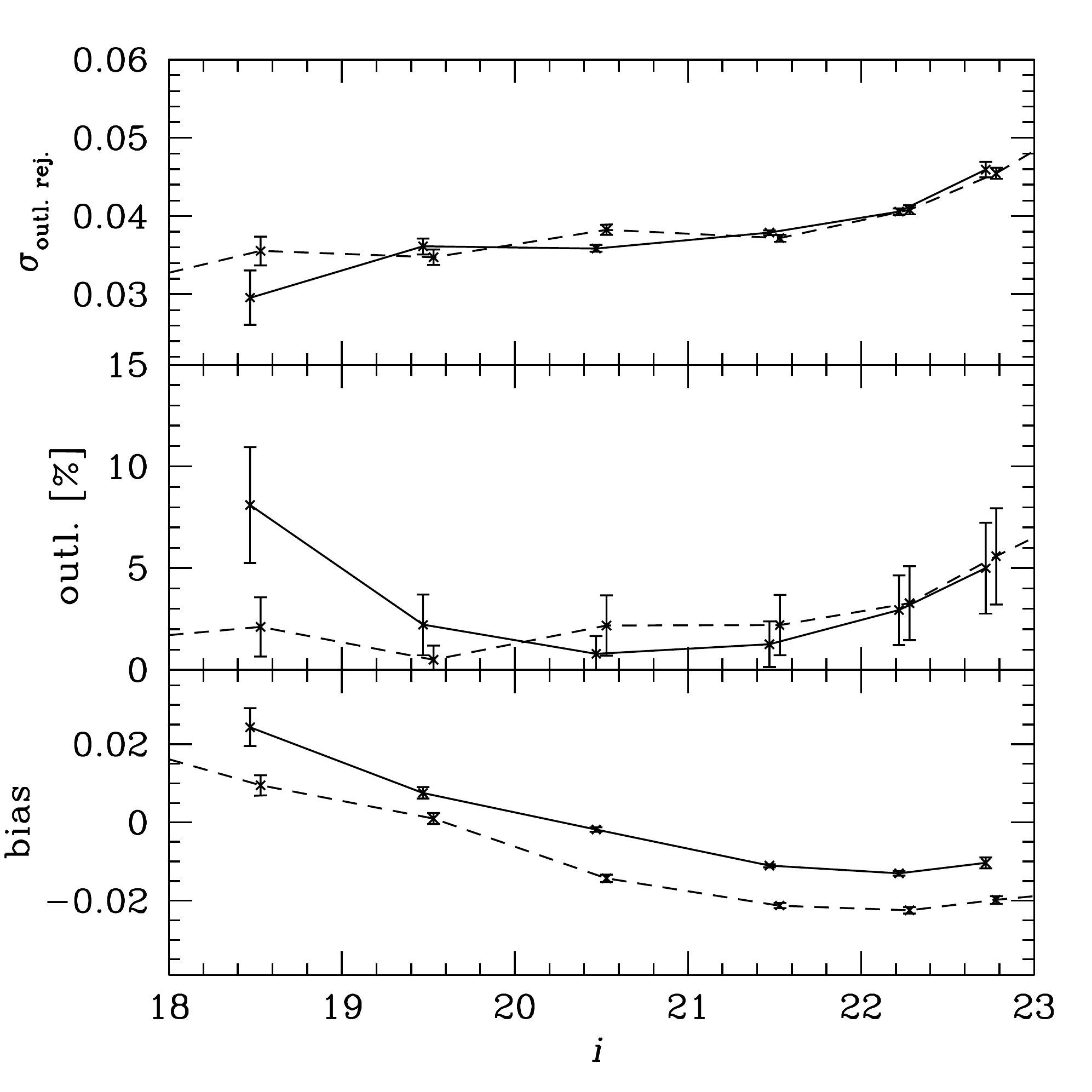}
\caption{Photo-$z$ statistics as a function of magnitude. The top panel shows
the photo-$z$ scatter after outliers were rejected, the middle panel shows the
outlier rate, and the bottom panel shows the bias (outliers included;
positive means photo-$z$’s overestimate the spec-$z$’s). Errors are purely
Poissonian. Note that the errors between magnitude bins are correlated.
The solid curve shows statistics for the analysis of this article.
For comparison we also show corresponding measurement from \citet{hek12}
(dashed curve).}
\label{fig:zstats_mag}
\end{figure}

\begin{figure}
\includegraphics[width=0.95\columnwidth]{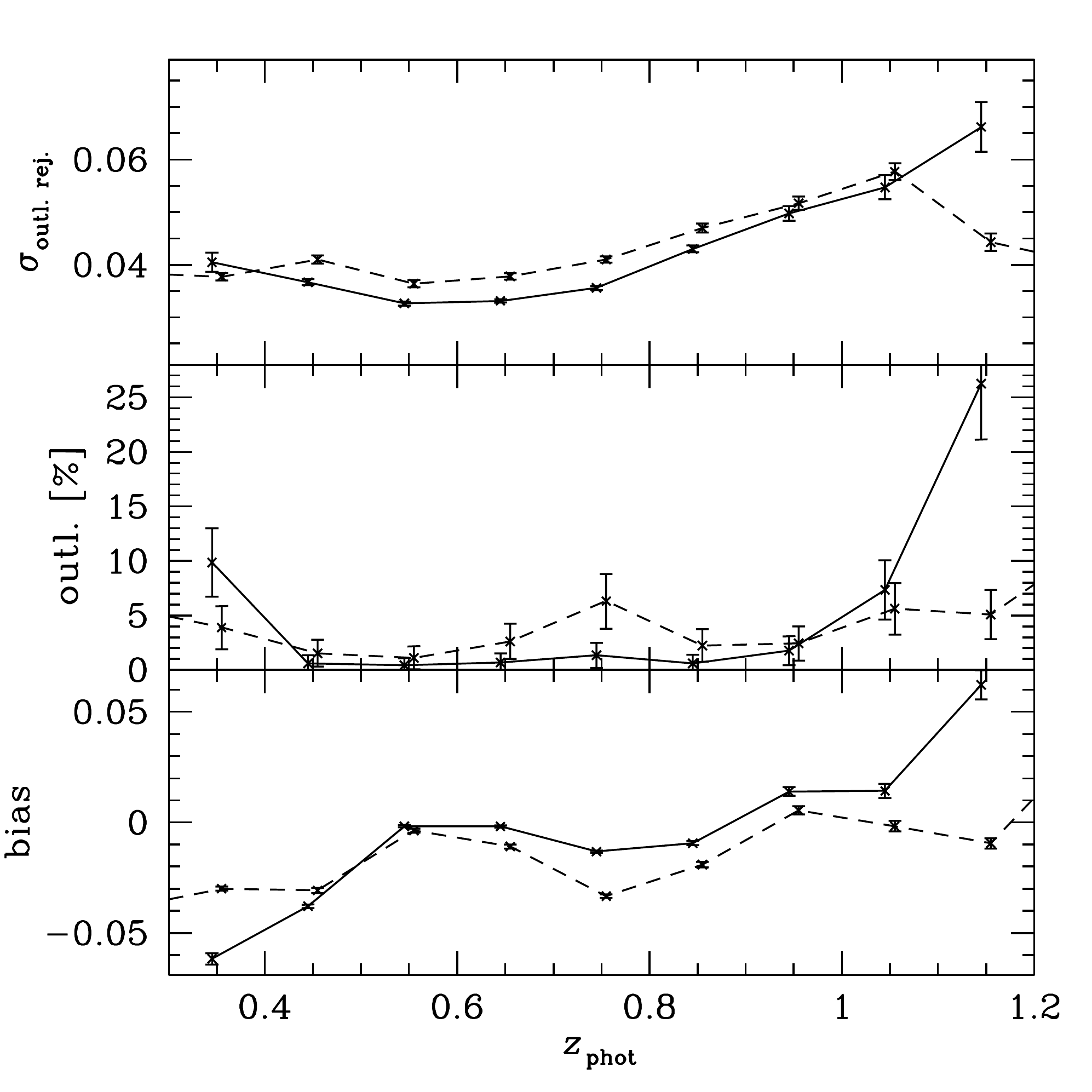}
\caption{Similar to \figref{fig:zstats_mag} but here statistics are a
  function of photo-$z$. We only plot the redshift interval where
  VIPERS yields a sufficient number of spec-$z$. The solid curve shows
  statistics for the analysis of this article.  For comparison we also
  show corresponding measurement from \citet{hek12} (dashed curve).}
\label{fig:zstats_z}
\end{figure}

Figure~\ref{fig:zz} shows a direct comparison of the \CFHTLenS
photo-$z$ versus \VIPERS spec-$z$ of 18\,995 objects. Note that the
\VIPERS spec-$z$ catalogue is pre-selected by colour, targeting mostly
objects in the range $0.5\la z\la1.2$ down to $i'\approx 22.5$.
We estimate photo-$z$ statistics (scatter, outlier rate,
bias, and completeness) as a function of $i'$-band magnitude and
redshift in the same way as described in \cite{hek12}. The results are
shown in Figs.~\ref{fig:zstats_mag} and \ref{fig:zstats_z}. Comparing
to the performance of the \CFHTLenS photo-$z$ vs. \VVDS/\DEEPT/\Sloan
spec-$z$ we do not find any significant differences in the magnitude
range ($i'\la22.5$) and redshift range ($0.5\la z\la1.2$), where
\VIPERS spec-$z$ are available.

This test suggests that the photo-$z$ accuracy (and hence also the
photometry) is stable over the survey area, beyond the fields that
could be tested with the original spec-$z$ catalogues. Having such a
successful blind test - a posteriori - is a strong argument for the
stability of our global photometry, and confirms that the photo-$z$
statistics presented in \cite{hek12} can be assumed for the whole
survey with a greater degree of confidence.
\subsection{Galaxy Correlation Functions on Large Angular Scales}
As a further test for the photometric homogeneity of our data beyond
individual pointings we investigate the galaxy correlation function
out to large angular scales.  The behaviour of the large-scale
galaxy angular correlation function, $w(\theta)$, is a sensitive
diagnostic test of large-scale systematic photometric gradients in an
imaging dataset.  Such photometric gradients would cause systematic
density variations in a source sample selected above a given flux
threshold. Since the random comparison datasets used to estimate
$w(\theta)$ are generated assuming a uniform source density, any such
gradient will result in an excess of signal in the large-scale
$w(\theta)$ such that it does not asymptote to zero.
In contrast to the tests described above, we here use our
patch-wide science object catalogues described in 
\citet{hek12} and \appendixref{sec:cfhtlenscatalogues}.
We use all galaxies down to $i' = 22$, which results
in the following sample sizes: $656\,998$ galaxies in \Wone,
$217\,359$ in \Wtwo, $483\,333$ in \Wthree and
$189\,209$ in \Wfour. The random comparison catalogues in each
patch have four times the corresponding object count.

We measure $w(\theta)$ for all four CFHTLenS regions in 30
logarithmic angular bins between 0.003 and 3 degrees with the
\citet{las93} estimator.  We restrict the sample to objects with
star-galaxy classifier $\mbox{\texttt{star\_flag}}=0$ and
$\mbox{\texttt{MASK}}=0$
(see \appendixref{sec:cfhtlenscatalogues}), and consider three
magnitude thresholds $i' < (20,21,22)$.  The integral constraint
correction is applied to the correlation functions.  We determine
the errors in the measurement using jack-knife re-sampling.  The
jack-knife samples are extended across all four regions such that
each sample has a characteristic size of 3 degrees; 54 jack-knife
samples are used in total.  We note that the measurements in the
different \CFHTLenS regions produce consistent results within the
expectation of cosmic variance, with the dispersion between the
regions becoming minuscule at small angular scales.

\begin{figure}
\includegraphics[width=0.95\columnwidth]{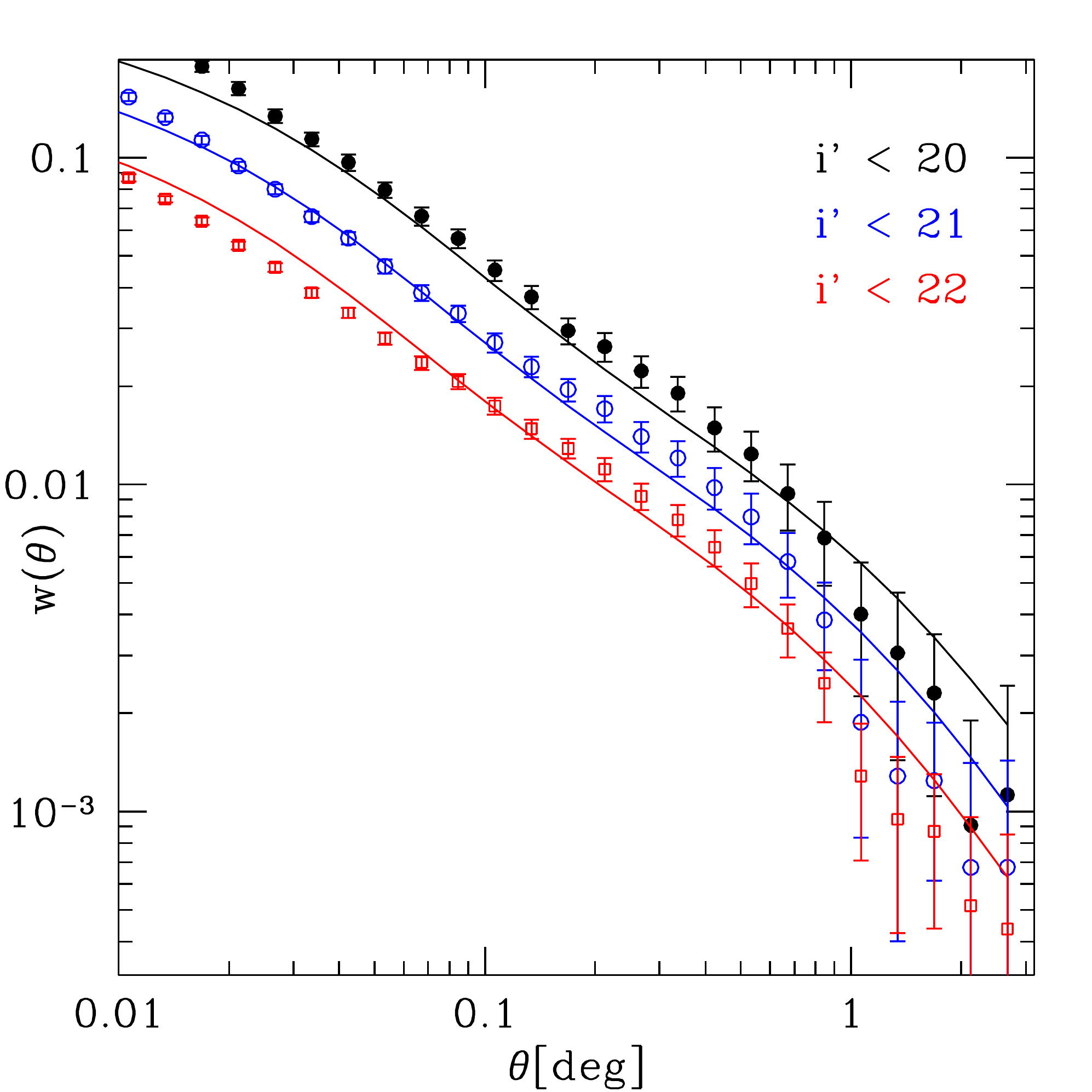}
\caption{Combined angular galaxy correlation functions for \CFHTLenS
patches \Wone to \Wfour; see text for further details.}
\label{fig:correlfunc}
\end{figure}

The combined correlation function measurements of the four patches are
plotted in \figref{fig:correlfunc} and compared to the predictions of
a $\Lambda$CDM cosmological model following \citet{spj03}.  This
prediction is generated from a CAMB \citep[see][]{lcl00} + halofit
non-linear power spectrum (produced using cosmological parameters
consistent with the latest CMB measurements), combined with galaxy
redshift distributions produced by stacking the photometric redshift
probability distributions at each magnitude threshold, and assuming a
linear galaxy bias factor $b \sim 1.2$.  The measurements are
consistent with the model at large scales, and tend to zero there,
revealing no evidence for systematic photometric gradients in the
sample.  The model is not expected to be a good match to the data at
scales below $1 \mbox{Mpc}/h_{72}$\footnote{$1 \mbox{Mpc}/h_{72}$
  subtends about 0.04 degree at the median redshift ($z_{\rm
    med}\approx 0.7$) of \CFHTLenS.}, where non-linear and halo model
effects become important.

We stress that we present this analysis primarily to further strengthen
confidence in the integrity of our photometric catalogues. We do not
want to present an in-depth investigation of the angular galaxy correlation
function or to interpret it scientifically. This will be done in
Bonnett et al. (in preparation).

%
\section{Released Data Products}
In the spirit of the \CFHTLS we make all data used for scientific
exploitation by the \CFHTLenS team available to the astronomical
community. The released data package includes:
\begin{enumerate}
\item The co-added \CFHTLenS pixel data products consisting of primary
  science data, weight- and flag-maps, sum frames and image masks. All
  these products are introduced and described
  in \sectionref{sec:coaddmask}.  Important details for potential
  users are provided in \appendixref{sec:cfhtlensimaging}.
\item The \CFHTLenS source catalogues with all relevant photo-$z$ and
  lensing/shear quantities. The creation of these catalogues is
  described in \citet{hek12} and \citet{mhk12}. The catalogue entries
  are described in \appendixref{sec:cfhtlenscatalogues}.
\end{enumerate}
The data are made available by \CADC through a web interface and can
be found at:
\url{http://www.cadc-ccda.hia-iha.nrc-cnrc.gc.ca/community/CFHTLens/query.html}.
The interface allows users to retrieve image pixel data on a
pointing/filter basis.  The catalogues can be accessed with a
sky-coordinate query form with filter options on all catalogue
entries.
\section{Conclusions}
We have presented the \CFHTLenS data products that originate from the
\CFHTLS-Wide survey. \CFHTLS-Wide was specifically designed as a weak
lensing survey providing deep, high quality optical data in five
passbands. Prior to the scientific exploitation of the data, the
\CFHTLenS collaboration had the objective to develop and to
thoroughly verify all necessary algorithms and tools in order to fully
exploit the survey.  This development includes numerous refinements to
existing data processing techniques, in particular an optimal treatment
in the astrometric and photometric calibration phase. Another
important upgrade of our analysis was to develop an algorithm to
nearly automatically perform the important image masking task. Hitherto,
it has mainly been performed manually. It is important to stress that
specific, high-precision scientific applications such as our weak
lensing analyses generally require very specific data processing
steps. These often tend to be in conflict with a
\emph{general-purpose} data set which needs to fulfil the requirements
of diverse scientific applications. Where necessary, our data
processing was heavily specialised to analyse small and faint
background sources that are essential for all weak lensing
studies. This affects for instance our sky-background subtraction
which aims for a \emph{local} sky-background as flat as possible on
small angular scales.  Furthermore, our treatment of cosmic rays has been
optimised for a robust identification of cosmic ray hits on the basis
of individual images.  This was crucial for the \lensfit shear
pipeline which entirely operates on single frames instead of the
co-added images. As described in
\sectionref{sec:cosmicrays}, our current implementation leads to a
strong incompleteness of stellar counts at faint magnitudes. For this
reason the \CFHTLenS data is complementary to
other publicly released versions of the \CFHTLS-Wide
Survey\footnote{The \CFHTLS releases of \Terapix (see
  \url{terapix.iap.fr}) and the \MegaPipe effort \citep[see][]{gwy08}
  can be obtained at
  \url{http://www3.cadc-ccda.hia-iha.nrc-cnrc.gc.ca/cfht/cfhtls_info.html}.}.

We have demonstrated that we are able to produce a homogeneous and
high-quality data set suitable for weak lensing studies with
photometric redshift estimates.  Our external astrometric accuracy
with respect to \Sloan data is around 60-70 mas, the internal
alignment in all filters is around 30 mas.  Our average photometric
calibration shows a dispersion with respect to \Sloan on the order of
0.01 to 0.03 mag for $g'$, $r'$, $i'$ and $z'$ and about 0.04 mag for
$u^*$.  We show in \citet{hvm12}, \citet{mhk12} and \citet{hek12} that
our data have the necessary quality to fully exploit the scientific
potential of a 154 deg$^2$ weak lensing survey.

The newly available \SDSSRE data which, covering almost the complete
\CFHTLenS area, will allow us to further refine our algorithms and
procedures in the future, especially increasing the quality
of our photometry. This will be particularly useful in preparation
for the next generation of weak lensing surveys that will cover
substantial parts of the sky, such as the 1500 deg$^2$ Kilo-Degree
Survey\footnote{\url{http://kids.strw.leidenuniv.nl/}}
\citep[][]{jkk12} or the 5000 deg$^2$ Dark Energy
Survey\footnote{\url{http://www.darkenergysurvey.org/}}
\citep[][]{mab12}.  For these surveys the accuracy of current
algorithms certainly needs to be further improved to exploit their
full scientific potential and to not be dominated by residual
systematics.

In the hope that we will trigger a variety of new developments
and follow-up studies with the \CFHTLenS products, we make the complete
data set, consisting of pixel data and object catalogues with all
relevant lensing and photo-$z$ quantities,
publicly available via \CADC. \\

\vspace*{5mm}
\noindent
{\large \bf ACKNOWLEDGEMENTS}\\
We thank the \VIPERS collaboration, for a fruitful data exchange and
for providing us with unpublished spectroscopic redshifts. We also
thank TERAPIX for the individual exposures quality assessment and
validation during the \CFHTLS data acquisition period, and Emmanuel
Bertin for developing key software modules used in this
study. \CFHTLenS data processing was made possible thanks to
significant computing support from the NSERC Research Tools and
Instruments grant program. We thank the \CFHT staff for successfully
conducting the \CFHTLS observations and in particular Jean-Charles
Cuillandre and Eugene Magnier for the continuous improvement of the
instrument calibration and the \Elixir detrended data that we used.

This work is based on observations obtained with \MegaPrime/\MegaCam,
a joint project of CFHT and CEA/DAPNIA, at the Canada-France-Hawaii
Telescope (\CFHT) which is operated by the National Research Council
(NRC) of Canada, the Institut National des Sciences de l'Univers of
the Centre National de la Recherche Scientifique (CNRS) of France, and
the University of Hawaii. This research used the facilities of the
Canadian Astronomy Data Centre operated by the National Research
Council of Canada with the support of the Canadian Space Agency.

TE is supported by the Deutsche Forschungsgemeinschaft through project ER 327/3-1 and the Transregional Collaborative Research Centre TR 33 - "The Dark Universe".
HH is supported by the Marie Curie IOF 252760, a CITA National Fellowship,
and the DFG grant Hi 1495/2-1.
LVW acknowledges support from the Natural Sciences and Engineering Research Council of Canada (NSERC) and the Canadian Institute for Advanced Research (CIfAR, Cosmology and Gravity program).
CH acknowledges support from the European Research Council under the EC FP7 grant number 240185.
H. Hoekstra acknowledges support from  Marie Curie IRG grant 230924, the Netherlands Organisation for Scientific Research (NWO) grant number 639.042.814 and from the European Research Council under the EC FP7 grant number 279396.
TDK acknowledges support from a Royal Society University Research Fellowship.
YM acknowledges support from CNRS/INSU (Institut National des Sciences de l'Univers) and the Programme National Galaxies et Cosmologie (PNCG).
CB is supported by the Spanish Science MinistryAYA2009-13936 Consolider-Ingenio CSD2007-00060, project2009SGR1398 from Generalitat de Catalunya and by the the European Commission’s Marie Curie Initial Training Network CosmoComp (PITN-GA-2009-238356).
LF acknowledges support from NSFC grants 11103012 \& 10878003, Innovation Program 12ZZ134 and Chen Guang project 10CG46 of SMEC, and STCSM grant 11290706600.
KH acknowledges support from a doctoral fellowship awarded by the Research council of Norway, project number 177254/V30.
MJH acknowledges support from the Natural Sciences and Engineering Research Council of Canada (NSERC).
MK is supported in parts by the DFG cluster of excellence `Origin and Structure of the Universe'. 
BR acknowledges support from the European Research Council in the form of a Starting Grant with number 24067. 
TS acknowledges support from NSF through grant AST-0444059-001, SAO through grant GO0-11147A, and NWO.
ES acknowledges support from the Netherlands Organisation for Scientific Research (NWO) grant number 639.042.814 and support from the European Research Council under the EC FP7 grant number 279396. 
PS is supported by the Deutsche Forschungsgemeinschaft through the Transregional Collaborative Research Centre TR 33 - "The Dark Universe".
MS acknowledges support from the Netherlands Organization for Scientific Research (NWO).
MV acknowledges support from the Netherlands Organization for Scientific Research (NWO) and from the Beecroft Institute for Particle Astrophysics and Cosmology.

{\small Author Contributions: All authors contributed to the
  development and writing of this paper.  The authorship list reflects
  the lead authors of this paper (TE, HH, LM, LVW, CH) followed by two
  alphabetical groups.  The first alphabetical group includes key
  contributers to the science analysis and interpretation in this
  paper, the founding core team and those whose long-term significant
  effort produced the final \CFHTLenS data product.  The second group
  covers members of the \CFHTLenS team who made a contribution to the
  project and/or this paper. CH and LVW co-led the \CFHTLenS
  collaboration.}

\bibliographystyle{mn2e}
\bibliography{CFHTLenS}

\label{lastpage}

\appendix
\section{\CFHTLenS pointing quality information}
\label{app:cfhtlensqual}
In \tabref{tab:CFHTLenSquality} we provide detailed information about the
characteristics of all \CFHTLenS fields. It contains the effective
area of each field after image masking ($\mbox{\texttt{MASK}}=0$ areas;
see \sectionref{sec:coaddmask}), the number of individual images
contributing to each stack, the total exposure time, the limiting
magnitude as defined in
\sectionref{sec:cfhtlens}, magnitude comparisons with \Sloan as described in
\sectionref{sec:astromphotomqual}, the measured image seeing and
special comments.  We note again that the magnitude comparison is
based on object catalogues extracted from each individual \CFHTLenS
pointing. The magnitude used for the comparison is the \SExtractor
quantity \texttt{MAG\_AUTO} for all filters. We do not show
direct magnitude comparisons with the \CFHTLenS catalogues described
in \appendixref{sec:cfhtlenscatalogues}. We have verified that differences
of the \texttt{MAG\_$x$} with $x\in \{u, g, r, i , y, z\}$ quantity
in the \CFHTLenS catalogues are close to the values quoted here.

In the comments field of \tabref{tab:CFHTLenSquality} we use the following
abbreviations:
\begin{itemize}
\item \textbf{no ch. XX:} The stack contains no data around chip
  position(s) XX. We number the \MegaPrime mosaic chip from left to
  right and from bottom to top. The lower left (east-south) chip has
  number 1, the lower right (west-south) chip number 9 and the
  upper-right (west-north) chip number 36. Note that this labeling
  scheme differs from that used at \CFHT.
\item \textbf{obv. st. break:} The stellar locus in a size
  vs. magnitude diagram shows a clear \emph{stellar break} as
  discussed in
  \sectionref{sec:cosmicrays}. The judgement was done on a subjective
  basis by visually inspecting FLUX\_RADIUS vs. MAG\_AUTO diagrams for
  all pointings and filters.
\item \textbf{WL pass:} The field passes the \CFHTLenS \emph{Weak
    Lensing Field Selection} as described in Sect. 4.2 of
  \citet{hvm12}.
\end{itemize}
\onecolumn
\begin{longtable}{llrrrrrrp{2.5cm}}
\caption{\label{tab:CFHTLenSquality}\CFHTLenS data quality overview:
Magnitude offsets are given as
$\Delta m = m_{\rm CFHTLenS}-m_{\rm SDSS}$. See the text for more details.} \\
\hline\hline
\multicolumn{1}{c}{Field/Area} & \multicolumn{1}{c}{Filter} & \multicolumn{1}{c}{N} & \multicolumn{1}{c}{expos. time} & \multicolumn{1}{c}{$m_{\rm lim}$} & \multicolumn{1}{c}{Sloan} & \multicolumn{1}{c}{seeing} & \multicolumn{1}{c}{comments} \\
\multicolumn{1}{c}{[sq. deg.]} & & & \multicolumn{1}{c}{[s]} & \multicolumn{1}{c}{[AB mag]} & \multicolumn{1}{c}{$\Delta m \times 100$} & \multicolumn{1}{c}{[$''$]} & \\
\hline
\endfirsthead
\hline\hline
\multicolumn{1}{c}{Field/Area} & \multicolumn{1}{c}{Filter} & \multicolumn{1}{c}{N} & \multicolumn{1}{c}{expos. time} & \multicolumn{1}{c}{$m_{\rm lim}$} & \multicolumn{1}{c}{Sloan} & \multicolumn{1}{c}{seeing} & \multicolumn{1}{c}{comments} \\
\multicolumn{1}{c}{[sq. deg.]} & & & \multicolumn{1}{c}{[s]} & \multicolumn{1}{c}{[AB mag]} & \multicolumn{1}{c}{$\Delta m \times 100$} & \multicolumn{1}{c}{[$''$]} & \\
\hline
\endhead
\hline
\endfoot
\texttt{W1m0m0} & $u^*$ & 5 & 3000.26 & 25.17 & $-6.8 \pm 4.0$ & 0.78 &  \\
(0.76) & $g'$ & 5 & 2500.37 & 25.44 & $-1.7 \pm 2.4$ & 0.78 &  \\
 & $r'$ & 4 & 2000.34 & 25.00 & $-0.5 \pm 3.1$ & 0.64 &  \\
 & $i'$ & 8 & 4920.69 & 24.54 & $-0.3 \pm 3.3$ & 0.63 & WL pass \\
 & $z'$ & 6 & 3600.46 & 23.17 & $-2.9 \pm 4.9$ & 0.92 &  \\
\hline
\texttt{W1m0m1} & $u^*$ & 5 & 3000.21 & 25.39 & $-1.8 \pm 3.6$ & 1.00 &  \\
(0.80) & $g'$ & 5 & 2500.33 & 25.71 & $-0.9 \pm 2.2$ & 0.86 &  \\
 & $r'$ & 4 & 2000.28 & 24.95 & $-0.0 \pm 2.5$ & 0.71 &  \\
\hline
\hline
\end{longtable}

\twocolumn

\section{\CFHTLenS Imaging Products}
\label{sec:cfhtlensimaging}
The \CFHTLenS imaging data release contains the essential products
after the co-addition and masking phase
(see \sectionref{sec:coaddmask}). The package consists of: (1) The
primary science pixel data from all pointings for all available
filters. (2) Weight maps characterising the sky-noise properties in
each pixel of the primary science data. The weights contain
\emph{relative} weights of the pixels in the science data. The
\SExtractor \texttt{WEIGHT\_TYPE} to use for object analysis is
\texttt{MAP\_WEIGHT}.  (3) A \texttt{flag} image which has a $0$ where
the weight is unequal to zero and a $1$ where the weight is zero,
i.e. a $1$ indicates a pixel in the co-added science image to which
none of the single frames contributed.  (4) \emph{sum} images are
integer pixel data whose pixel value correspond to the number of input
images contributing to the corresponding pixel of the science
data. (5) \emph{mask} images encoding the results of our masking
procedures. Note that we do not officially release any products from
the eight \Wtwo pointings with incomplete colour coverage; see
\figref{fig:cfhtlenslayout}. The \CFHTLenS team only processed these
pointings up to the image co-addition phase but did not create object
catalogues for these fields. Interested readers can obtain imaging
data products of these fields (except mask files) by request to the
authors.

All data are self-contained to easily allow further processing. All
necessary information to relate image pixel positions to sky
coordinates and flux values to apparent magnitudes is provided in the
form of FITS image header keywords. Astrometric header items follow
standard World Coordinate System descriptions as described in
\citet{grc02}. The essential header keywords to extract photometric
information are summarised in \tabref{tab:photheaders}.

To reject obviously problematic sources from an object catalogue
extracted from \CFHTLenS images, everything that contains pixels that
have a 1 in the \texttt{flag} should be removed. A much more
sophisticated and fine-tuned catalogue cleaning can be done with our
\texttt{mask} files. It encodes areas from our masking procedures
(see \sectionref{sec:coaddmask}) as well as information from all the
\texttt{flag} images of all filters.  The coding of the pixel
values in this image is given in
\tabref{tab:maskvalues}. The primary reference of our masking
procedures is the lensing band, i.e. the $i'$-band or $y'$-band
observation.  In particular, features not common to all filters
(e.g. asteroid tracks) are ensured to be masked only in these
passbands.  We first mask stars brighter than $m_{\rm GSC}<11$\footnote{Objects
  that need to be masked are identified primarily with the Guide Star
  Catalogues 1 and 2 \citep[see e.g.][]{llm08}.} with a wide mask that
encompasses the stellar halo and prominent diffraction spikes.  We
empirically determined that for our \CFHTLenS observations stars with $m_{\rm GSC} <
10.35$ should be masked in any case while many stellar haloes in the
range $11.0\leq m_{\rm GSC}\leq 10.35$ are only barely visible. In obvious
cases the corresponding mask was removed during our manual pass
through all image masks. Remaining stars down to $m_{\rm GSC}<17.5$ are
surrounded with a template that is scaled with magnitude. In addition
we independently mask areas for the four filters $u^*$, $g'$, $r'$ and $i'$
whose object density distribution differs significantly from the
mean of the one square degree pointing. We found this to effectively
catch areas around large extended objects that we want to exclude in
our shear/lensing experiments. Rich galaxy clusters that have been
masked by this procedure were again unmasked during the manual
verification phase. The precise procedures to obtain the masks are
described in \citet{ehl09}.  All science analyses of the \CFHTLenS team is
performed with sources having a mask value of $\leq 1$. Details are
given in the corresponding science articles. When using \SExtractor the
flagging or masking information can be straightforwardly transfered to an
object catalogue by using the corresponding images as \emph{external
  flags}.

We note that we do not release sky-subtracted single frame data
products for the lensing bands. These data form the basis for our
shear analyses with \lensfit; \citep[see][]{mhk12}.  The data volume
of these products is very large and they are of interest for
a few groups only.  They can be obtained by request to the authors.
The same applies for the PSF homogenised versions of the co-added
images which were used to estimate object colours for our photo-$z$
estimates.
\begin{table}
  \caption{Description of important \CFHTLenS FITS Image Header Keywords}
\label{tab:photheaders}      
\centering          
\begin{tabular}{ll}    
\hline\hline       
Keyword & Description \\
\hline
\texttt{TEXPTIME} & total exposure time in seconds \\
\texttt{EXPTIME} & \emph{effective} exposure time. This is always 1$s$ for
\CFHTLenS \\
 & data; the pixel unit of all \CFHTLenS images is ADU/s \\
\texttt{MAGZP} &  magnitude zeropoint; apparent object AB magnitudes \\ 
 & need to be estimated via: \\ 
 & $mag = \mbox{MAGZP} - 2.5\log(\mbox{object counts})$ \\
\texttt{GAIN} & The effective median gain of the exposure. \\
 & To obtain meaningful magnitude error estimates within \\
 & \SExtractor the \texttt{GAIN} configuration parameter \\
 & needs to be set the the \texttt{GAIN} header value \\
\texttt{SEEING} & measured mean image seeing for the science image. \\
 & Put this value into the \texttt{SEEING\_FWHM} \SExtractor \\
 & parameter to obtain a meaningful \SExtractor \\
 & star/galaxy separation. \\
\hline                  
\end{tabular}
\end{table}
\begin{table}
  \caption{\label{tab:maskvalues}Description of values in \CFHTLenS
    masking data. Note that an actual pixel in a mask can be a sum
    of listed values; see text for further details.}
\centering          
\begin{tabular}{ll}    
\hline\hline       
\texttt{mask} value & Description \\
\hline
1 & large masks around stars and stellar haloes for objects \\
  & with $10.35 \leq m_{\rm GSC} \leq 11.00$. For a less conservative masking you \\
  & can consider using sources falling within these masks \\
2 & large masks around stars and stellar haloes for objects \\
  & with $m_{\rm GSC} < 10.35$ \\
4 & masks around asteroid trails in the lensing band \\
8 & $g'$-band mask around areas of significant object \\
  & overdensities and gradients in the object density \\
  & distribution \\
16 & $r'/i'/y'$-band mask around areas of significant object \\
   & overdensities and gradients in the object density \\
   & distribution \\
32 & $u^*$-band mask around areas of significant object \\
   & overdensities and gradients in the object density \\
   & distribution \\
64 & masks around bright stellar sources \\
128 & pixels flagged in the $i'/y'$-band \\
256 & pixels flagged in the $u^*$-band \\
512 & pixels flagged in the $g'$-band \\
1024 & pixels flagged in the $r'$-band \\
2048 & pixels flagged in the $z'$-band \\
8192 & the area is outside the \CFHTLenS catalogue of the \\
 & pointing (see \sectionref{sec:cfhtlenscatalogues}) \\
\hline                  
\end{tabular}
\end{table}
\section{\CFHTLenS Catalogue products}
\label{sec:cfhtlenscatalogues}
The \CADC data release interface\footnote{please visit
\url{http://www.cadc-ccda.hia-iha.nrc-cnrc.gc.ca/community/CFHTLens/query.html}}
allows users to query and retrieve the \CFHTLenS catalogue that our
team is using for all analyses. In this section we briefly summarise
the catalogue creation procedures and we explain all relevant
catalogue entries. 
 
The catalogue are created starting from the co-added \CFHTLenS
images (see \sectionref{sec:coaddmask}).  In short we perform the
following steps to create catalogues on a pointing basis:
\begin{enumerate}
\item From an initial \SExtractor source list we extracted catalogues of
  stellar sources for each pointing
  in the lensing band. To have a high-confidence catalogue for the
  crucial steps of PSF mapping and PSF homogenisation this step
  was performed manually with the help of stellar locus diagrams.
\item The PSFs of each \CFHTLenS pointing were gaussianised to the
  seeing of the worst image quality amongst the five filters.
  This step yields new versions of the co-added data which are subsequently
  used to estimate robust galaxy colours \citep[][]{hek12}.
\item \SExtractor is run in dual image mode six times. The detection
  image is always the unconvolved lensing band image and the
  measurement images are the Gaussianised images in the five bands and
  - in the sixth run - the unconvolved lensing band image. This last
  run is performed to obtain total magnitudes \citep[\SExtractor
  quantity MAG\_AUTO][]{kro80} in the lensing band, whereas the first
  five runs yield accurate colours based on isophotal magnitudes.
\item We add a position dependent estimate for the limiting magnitude
  to each object. This is done with the help of \SExtractor
  RMS check images, which contain an estimate of the sky-background
  variation on each pixel position. Limiting magnitudes are estimated
  within the seeing disk as described in \citet{hek12}.
\item Galactic extinction values on each object position are added
  based on the \cite{sfd98} dust maps.
\item The estimated total magnitudes in the lensing band (see above)
  are combined with the colour estimates, the limiting magnitudes (to
  decide whether an object is detected in a given band), and the
  extinction values to yield estimates of the total magnitudes in the
  other bands. This procedure assumes that there are no colour
  gradients in the objects. For galaxies with colour gradients the
  total magnitudes in the $u^*g'r'z'$-bands might be biased and only
  the lensing band total magnitudes are reliable.
\item A mask column based on the final, eye-balled and modified masks
  (see \sectionref{sec:coaddmask}) is added to the object entries.
\item We use the Bayesian Photometric Redshift Code
  \citep[\BPZ;][]{ben00} to estimate photo-$z$. Instead of the standard
  template set provided by \texttt{BPZ} we use
  a recalibrated one described in \cite{cap04}.
\item Absolute rest-frame magnitudes in the \MegaPrime filters as well
  as stellar masses (see Velander et al., submitted to MNRAS, for details)
  are added based on the \BPZ photo-$z$ estimate and a best-fit
  template from the \cite{brc03} library. This step is performed
  keeping the redshift fixed using the \LePhare code \citep{amv02} and
  the \cite{isl10} technique.
\item From each \CFHTLenS pointing catalogue we cut away overlap regions
  with neighbouring pointings. This avoids issues with multiple entries
  for a specific source when the pointing based catalogues are
  merged to a \emph{patch-wide} source list. Areas that are cut out
  in this way are specifically marked in our mask files with a value
  of 8192; see \tabref{tab:maskvalues}.
\end{enumerate}
The last step concludes the estimation of all photometry related
quantities in the \CFHTLenS catalogues. Important additional details
of the photometric catalogue creation can be found in \citet{hek12}.

The star and galaxy catalogues were then passed to the \lensfit shear
analysis of the individual exposures as described by \citet{mhk12} and
\citet{hvm12}.  The multiplicative and additive shear calibration
factors described by \citet{mhk12}, eq.~(14), and \citet{hvm12}
, eq.~(19), may be calculated from the quantities \verb1scalelength1
and \verb1SNratio1 given below.

\tabref{tab:catalogue_cols} lists all relevant catalogue entries that
can be retrieved from the \CADC interface. We list the column name, a
short description, the software to estimate the quantity and the
units.  Most quantities refer to the lensing band that served as the
detection image. If a quantity relates to another band this is
indicated directly in the quantity names with an \verb1_x1 where
\texttt{x} is either [ugriyz].

In the following we give additional information on certain columns in
the catalogue:
\begin{itemize}
\item \verb1field1: The \CFHTLenS string identifier such as {W1m0m0}.

\item \verb1MASK1: The mask column as described in
  \tabref{tab:maskvalues}. If \verb1MASK1$>0$ the object centre lies
  within a mask. Objects with \verb1MASK1$\le 1$ can safely be used
  for most scientific purposes. Objects with \verb1MASK1$>1$ have been
  removed from the released catalogues.

\item \verb1T_B1: \BPZ spectral type. 1=CWW-Ell, 2=CWW-Sbc,
  3=CWW-Scd, 4=CWW-Im, 5=KIN-SB3, 6=KIN-SB2. Note that the templates
  are interpolated; hence fractional types occur.

\item \verb1NBPZ_FILT, NBPZ_FLAGFILT, NBPZ_NONDETFILT1: The number of
  filters in which an object has \emph{reliable photometry}
  (\verb1NBPZ_FILT1), i.e. magnitude errors $<1\mbox{mag}$ and objects
  brighter than the limiting magnitude; number of filters in which an
  object has formal magnitude errors of 1 mag or larger
  (\verb1NBPZ_FLAGFILT1); number of filters in which an object is
  fainter than the formal limiting magnitude
  (\verb1NBPZ_NONDETFILT1). If an object would fall into
  \verb1FLAGFILT1 as well as \verb1NONDETFILT1 it is listed under
  \verb1FLAGFILT1.  Magnitude errors refer to \verb1MAG_ISO1.

\item \verb1BPZ_FILT, BPZ_FLAGFILT, BPZ_NONDETFILT1: These keys
  contain a binary encoding to identify filters with problematic
  photometric properties for photo-$z$ estimation. Filter $u^*$ is
  assigned a '1', $g'$ = '2', $r'$ = '4',
  $i'$/$y'$ = '8' and $z'$ = '16'. The keys \verb1BPZ_FILT1,
  \verb1BPZ_FLAGFILT1 and \verb1BPZ_NONDETFILT1
  represent the sums of the filters fulfilling the criteria
  detailed for \verb1NBPZ_FILT1 etc.

\item \verb1PZ_full1: This is the full photometric redshift
  probability distribution $P(z)$ to $z=3.5$. There are 70
  columns sampling $P(z)$ at intervals of ${\rm d}z=0.05$. The first
  bin is centred at $z=0.025$. Note these 70 columns do not always
  sum to 1. There is a final bin not included in the catalogues with
  $z>3.5$ that, in a small number of cases, has non-zero
  probability. In \CFHTLenS analysis we set a hard prior of a
  zero probability past $z>3.5$, which corresponds to normalising each
  $P(z)$ to one. For future flexibility however we do not impose this
  normalisation on the catalogue, leaving it to the user to apply.

\item \verb1star_flag1: Stars and galaxies are separated using a
  combination of size, lensing band magnitude and colour
  information. For $i'<21$, all objects with size smaller than the PSF
  are classified as stars. For $i'>23$, all objects are classified as
  galaxies. In the range $21<i'<23$, a star is defined as size$<$PSF
  and $\chi^2_{\rm star}<2.0\chi^2_{\rm gal}$, with $\chi^2$
  the best fit $\chi^2$ from the galaxy and star libraries given by
  \LePhare.

\item \verb1MAG_LIM_[ugriyz]1: These are 1-$\sigma$ limiting
  magnitudes measured in a circular aperture with a diameter of
  $2\times$FWHM, where FWHM is the seeing in this band (see
  \texttt{SEEING} keyword in the image header).

\item \verb1weight1: The \lensfit inverse-variance weight to be used
  in the shear measurement for each galaxy as given by equation\,8 of
  \citet{mhk12}.

\item \verb1fitclass1: Object classification as returned by \lensfit.
  Possible classification values are:
\begin{center}
\begin{tabular}{rl}
0 & galaxy \\
1 & star \\
-1 & no fit attempted: no useable data \\ 
-2 & no fit attempted: blended or complex object \\ 
-3 & no fit attempted: miscellaneous reason \\ 
-4 & bad fit: $\chi^2$ exceeds critical value \\
\end{tabular}
\end{center}

\item \verb1scalelength1: \lensfit galaxy model scalelength.

\item \verb1bulge-fraction1: \lensfit galaxy model bulge fraction, $B/T$.
  The galaxy model disk fraction is $1-B/T$.

\item \verb1model-flux1: \lensfit galaxy model total flux, in
  calibrated CCD data units.

\item \verb1SNratio1: \lensfit signal-to-noise ratio of the object,
  measured within a limiting isophote 2-$\sigma$ above the noise.


\item \verb3PSF-e1, PSF-e23: \lensfit mean of the PSF ellipticity
  values measured on each exposure at the location of the galaxy.
  PSF ellipticities are derived
  from the PSF model at the location of each galaxy and are top-hat
  weighted with radius 8\,pixels.

\item \verb1PSF-Strehl-ratio1: mean of a set of `pseudo-Strehl ratio'
  values for the PSF model calculated on each exposure.  The pseudo-Strehl
  ratio is defined as the fraction of light in the PSF model that
  falls into the central pixel, and is a measure of the sharpness of
  the PSF.


\item \verb3e1, e23: \lensfit raw uncalibrated expectation values of
  galaxy ellipticity, from the marginalised galaxy ellipticity
  likelihood surface, to be used for shear measurement. We strongly
  urge the user not to use these raw uncalibrated ellipticity values
  blindly in a lensing analysis. First, any shear measurement must
  measure weighted averages using the lensfit weight. Secondly an
  additive \texttt{c2} correction must be applied to the \texttt{e2}
  component. This can be calculated from eq.~(19) of \citet{hvm12}
  from the qualities \texttt{scalelength} and \texttt{SNratio} noting
  that eq.~(19) is given in physical units (arcsec) whereas
  \texttt{scalelength} is given in pixel units.  One \MegaCam CCD
  pixel is $0\myarcsec 187$. Thirdly a multiplicative shear calibration
  correction must be applied following equations 15-17 of
  \citet{mhk12}. Note that it is incorrect to apply this
  multiplicative correction on an object by object basis. Instead this
  calibration correction must be applied as as an ensemble average
  (see Sect. 4.1 of \citealp{hvm12} for a summary of the required
  calibration corrections).  Finally, for any study that uses a shear
  two-point correlation function, only the fields that pass the
  systematics tests of \citet{hvm12} can be used. For other studies,
  such as galaxy-galaxy lensing or cluster studies we recommend that
  the measurement is made and compared for the full data set and the
  75\% of the data which passes the field selection. In the
  galaxy-galaxy lensing analysis of Velander et al.  (submitted to
  MNRAS) we find no difference between these two results.  We also
  note that \texttt{e2} is defined relative to a decreasing RA such
  that the user may need to multiply \texttt{e2} by $-1$ when defining
  angles in the RA/Dec reference frame (see Kilbinger et al.,
  submitted to MNRAS, for a discussion on calculating angles on a
  sphere).

\item \verb1n-exposures-used1: the number of individual exposures used
  by \lensfit for this galaxy.

\item \verb3PSF-e<1,2>-exp<i>3: the \lensfit PSF model ellipticity
  (top-hat weighted as above) on each exposure \verb1i1 at the
  location of the galaxy. An entry of $-99$ indicates that the object is
  either unobserved in the image (i.e a chip gap or, owing to the
  dithers, the object is off the edge of the image), or it indicates
  that the exposure does not exist. The majority of \CFHTLenS
  lensing band observations have 7
  exposures, but some have up to 15 hence there are 15 entries for
  each object.

\end{itemize}

\onecolumn
\begin{longtable}{llll}
\caption{\label{tab:catalogue_cols}\CFHTLenS Catalogue columns:
Quantities with a \_$x$ at the end of their name are present
for all available filters, i.e. $x\in \{u^*, g', r', i', y', z'\}$.}\\
  \hline
  \hline
  \multicolumn{1}{c}{column name} & \multicolumn{1}{c}{description} & 
  \multicolumn{1}{c}{programme} & \multicolumn{1}{c}{unit} \\
  \hline
  \endfirsthead
  \hline
  \hline
  \multicolumn{1}{c}{column name} & \multicolumn{1}{c}{description} & 
  \multicolumn{1}{c}{programme} & \multicolumn{1}{c}{unit} \\
  \hline
  \endhead
  \hline
  \endfoot
  \verb1id1 & Unique \CFHTLenS object identification ID. & \CADC & \\
  \verb1field1 & Name of the \CFHTLenS pointing. & \THELI & \\
  \verb1SeqNr1 & Running number within \CFHTLenS pointing & \SExtractor & \\
  \verb1Xpos1 & Centroid x-pixel position in the \CFHTLenS pointing & \SExtractor & pix\\
  \verb1Ypos1 & Centroid y-pixel position in the \CFHTLenS pointing & \SExtractor & pix\\
  \verb1ALPHA_J20001 & Centroid sky position right ascension & \SExtractor & deg\\
  \verb1DELTA_J20001 & Centroid sky position declination & \SExtractor & deg\\
  \verb1n_exposures_detec1 & Number of individual exposures contributing to the object's
  position. & \SExtractor & \\
  \verb1BackGr1 & Background counts at centroid position &
  \SExtractor &
  counts\\
  \verb1Level1 & Detection threshold above background. & \SExtractor & counts\\
  \verb1MU_MAX1 & Peak surface brightness above background &
  \SExtractor &
  mag $\cdot$ arcsec$^{-2}$\\
  \verb1MU_THRESHOLD1 & Detection threshold above background. &
  \SExtractor &
  mag $\cdot$ arcsec$^{-2}$\\
  \verb1MaxVal1 & Peak flux above background & \SExtractor & counts\\
  \verb1Flag1 & \SExtractor extraction flags. & \SExtractor &\\
  \verb1A_WORLD1 & Profile RMS along major axis. & \SExtractor & deg\\
  \verb1B_WORLD1 & Profile RMS along minor axis. & \SExtractor & deg\\
  \verb1THETA_J20001\footnote{In \SExtractor V2.4.6 the definition of the quantity
  THETA\_J2000 was changed and the sign flipped (Bertin, private communication). Because the
  \CFHTLenS catalogues were extracted with an older version users should be aware of this
  if they produce new source lists from the released \CFHTLenS data.} & 
    Position angle (east of north). & \SExtractor & deg\\
  \verb1ERRA_WORLD1 & World RMS position error along major axis. & \SExtractor&
   deg\\
  \verb1ERRB_WORLD1 & World RMS position error along minor axis. & \SExtractor &
   deg\\
  \verb1ERRTHETA_J20001 & J2000 error ellipse position angle &
   \SExtractor & deg\\
  \verb1FWHM_IMAGE1 & FWHM assuming a gaussian object profile & \SExtractor & pix\\
  \verb1FWHM_WORLD1 & FWHM assuming a gaussian object profile & \SExtractor & deg\\
  \verb1FLUX_RADIUS1 & Half-light radius. & \SExtractor & pixels\\
  \verb1CLASS_STAR1 & \SExtractor star-galaxy classifier & \SExtractor & \\
  \verb1MASK1 & CFHTLenS mask value at the object's position & \automask & \\
  \verb1ISOAREA_WORLD1 & Isophotal area above analysis threshold & 
   \SExtractor & deg$^2$\\
  \verb1NIMAFLAGS_ISO1 & Number of flagged pixels & \SExtractor & \\
  \verb1Z_B1 & BPZ redshift estimate; peak of posterior probability distribution & \BPZ & \\
  \verb1Z_B_MIN1 & Lower bound of the 95\% confidence interval of \verb1Z_B1 & \BPZ & \\
  \verb1Z_B_MAX1 & Upper bound of the 95\% confidence interval of \verb1Z_B1 & \BPZ & \\
  \verb1T_B1 & Spectral type corresponding to \verb1Z_B1 & \BPZ & \\
  \verb1ODDS1 & Empirical \texttt{ODDS} of \verb1Z_B1 & \BPZ & \\
  \verb1Z_ML1 & BPZ maximum likelihood redshift & \BPZ & \\
  \verb1T_ML1 & Spectral type corresponding to \verb1Z_ML1 & \BPZ & \\
  \verb1CHI_SQUARED_BPZ1 & $\chi^2$ value associated with \verb1Z_B1 & \BPZ & \\
  \verb1BPZ_FILT1 & Filters with good photometry (\texttt{BPZ}); bit-coded mask & \THELI & \\
  \verb1NBPZ_FILT1 & Number of filters with good photometry (\texttt{BPZ}) & \THELI & \\
  \verb1BPZ_NONDETFILT1 & Filters with faint photometry (not used in \texttt{BPZ}); bit-coded mask & \THELI & \\
  \verb1NBPZ_NONDETFILT1 & Number of filters with faint photometry (\texttt{BPZ}) & \THELI & \\
  \verb1BPZ_FLAGFILT1 & Filters with flagged photometry (\texttt{BPZ}); bit-coded mask & \THELI & \\
  \verb1NBPZ_FLAGFILT1 & Number of flagged filters (\texttt{BPZ}) & \THELI & \\
  \verb1LP_Mx1 & Absolute rest-frame magnitude in the $x$-band & \LePhare & mag\\
  \verb1star_flag1 & Star-galaxy separator ($0=$galaxy, $1=$star) & \LePhare & \\
  \verb2LP_log10_SM_MED2 & Logarithm of the stellar mass & \LePhare & $\log_{10}(M_\odot)$\\
  \verb2LP_log10_SM_INF2 & Lower bound of the logarithm of the stellar mass & \LePhare & $\log_{10}(M_\odot)$\\
  \verb2LP_log10_SM_SUP2 & Upper bound of the logarithm of the stellar mass & \LePhare & $\log_{10}(M_\odot)$\\
  \verb1PZ_full1 & Vector containing the posterior photo-z probability in steps of $\Delta_z=0.05$. & \BPZ & \\
  \verb1MAG_x1 & estimated total magnitude in the $x$-band & \SExtractor & mag\\
  \verb1MAGERR_x1 & Magnitude error in the $x$-band & \SExtractor & mag\\
  \verb1IMAFLAGS_ISO_x1 & $x$-band FLAG-image logically OR'ed flags values & \SExtractor & \\
  \verb1MAG_LIM_x1 & 1-$\sigma$ limiting magnitude in the $x$-band & \SExtractor & mag\\
  \verb1EXTINCTION_x1 & Galactic extinction in the $x$-band & \SExtractor & mag\\
  \verb1KRON_RADIUS1 & Scaling radius of the ellipse for magnitude measurements w.r.t. & \SExtractor & \\
 & \verb1A_WORLD1 and \verb1B_WORLD1 &  & \\
\hline
\verb1weight1 &            \lensfit weight  & \lensfit & \\
\verb1fitclass1 &          \lensfit fit class     & \lensfit & \\
\verb1scalelength1 &        \lensfit galaxy model scalelength & \lensfit & pix \\
\verb1bulge-fraction1 &     \lensfit galaxy model bulge-fraction & \lensfit & \\
\verb1model-flux1 &         \lensfit galaxy model flux & \lensfit & ADU \\
\verb1SNratio1 &            \lensfit data S/N ratio & \lensfit & \\
\verb3PSF-e1, PSF-e23 &     \lensfit PSF mean ellipticity components 1 and 2
& \lensfit & \\
\verb1PSF-Strehl-ratio1 &   \lensfit PSF pseudo-Strehl ratio & \lensfit & \\
\verb3e1, e23 &                  \lensfit galaxy e1, e2 expectation values & \lensfit & \\
\verb1n-exposures-used1 &   Number of exposures used in \lensfit measurement & \lensfit & \\
\verb3PSF-e<1,2>-exp<i>3 &  \lensfit PSF model e1, e2 on each exposure $i$ ($i=1,n$) & \lensfit & \\
\end{longtable}
\twocolumn

\end{document}